\documentclass[a4paper,11pt]{article}

\usepackage{geometry}										
\usepackage[english]{babel}									
\usepackage[utf8]{inputenc}									
\usepackage[T1]{fontenc}									

\usepackage{lmodern}										

\usepackage{amsmath}										
\usepackage{amstext}										
\usepackage{amsfonts}										
\usepackage{amssymb}										

\usepackage{bm}											
\usepackage{dsfont}										
\usepackage{units}										

\usepackage{textcomp}										

\usepackage{graphicx}										
\usepackage[svgnames]{xcolor}									
\usepackage{booktabs}										
\usepackage{multirow}										

\graphicspath{{./fig/}}

\newcommand{\cond}{\:\!|\:\!}									
\newcommand{\coeffM}{a}										
\newcommand{\basis}{\Psi}									
\newcommand{\dimParam}{M}																		
\newcommand{\dimData}{N}									
\newcommand{\maxDegree}{\gamma}									
\newcommand{\numTerms}{\Gamma}									
\newcommand{\functionOfInputs}{f}								

\newlength{\HYDROsubWidth}	
\newlength{\HYDROfigHeight}	
\newlength{\HYDROmapHeight}	
\newlength{\HYDROfigHeightNew}
\usepackage{authblk}										

\title{Principal component analysis and sparse polynomial chaos
  expansions for global sensitivity analysis and model calibration:
  application to urban drainage simulation}
\author[1]{Joseph B.\ Nagel\thanks{JosephBNagel@gmail.com}}
\author[2]{J\"{o}rg Rieckermann\thanks{joerg.rieckermann@eawag.ch}}
\author[1]{Bruno Sudret\thanks{sudret@ethz.ch}}
\affil[1]{Chair of Risk, Safety and Uncertainty Quantification, ETH Z\"{u}rich, Stefano-Franscini-Platz 5, 8093 Z\"{u}rich, Switzerland}
\affil[2]{Department Urban Water Management, Eawag, \"{U}berlandstrasse 133, 8600 D\"{u}bendorf, Switzerland}
\date{\today}

\usepackage{microtype}										

\usepackage{indentfirst}									

\usepackage{caption}										
\usepackage[labelformat=simple,subrefformat=simple,hypcap=true]{subcaption}			

\usepackage[numbers,sort&compress]{natbib}							
\usepackage{hyperref}
\hypersetup{
pdftitle = {Uncertainty quantification in urban drainage simulation: Fast surrogates for sensitivity analysis and model calibration},
pdfauthor = {Joseph B.\ Nagel, J\"{o}rg Rieckermann, Bruno Sudret},
pdfsubject = {Manuscript submitted to Environmental Modelling \& Software},
pdfkeywords = {Uncertainty quantification, surrogate modeling, polynomial chaos, dimension reduction, sensitivity analysis, Bayesian calibration},
colorlinks = false,										
}

\usepackage[capitalize]{cleveref}								
\creflabelformat{equation}{#2(#1)#3}								
\crefrangelabelformat{equation}{#3(#1)#4 to #5(#2)#6}						
\labelcrefformat{subequation}{#2(#1)#3}								
\labelcrefrangeformat{subequation}{#3(#1)#4 to #5(#2)#6}					

\newcommand{\revision}[1]{#1}

\begin{document}

\maketitle

\begin{abstract}
This paper presents an efficient surrogate modeling strategy for the uncertainty quantification and Bayesian calibration of a hydrological model. In particular, a process-based dynamical urban drainage simulator that predicts the discharge from a catchment area during a precipitation event is considered. The goal is to perform a global sensitivity analysis and to identify the unknown model parameters as well as the measurement and prediction errors. These objectives can only be achieved by cheapening the incurred computational costs, that is, lowering the number of necessary model runs. With this in mind, a regularity-exploiting metamodeling technique is proposed that enables fast uncertainty quantification. Principal component analysis is used for output dimensionality reduction and sparse polynomial chaos expansions are used for the emulation of the reduced outputs. Sensitivity measures such as the Sobol indices are obtained directly from the expansion coefficients. Bayesian inference via Markov chain Monte Carlo posterior sampling is drastically accelerated.
\end{abstract}

The quantification of uncertainty has become an integral aspect of
computational science and engineering in the last \revision{two decades}
\citep{Uncertainty:Smith2014,Uncertainty:Soize2017}.  Algorithmic
advances on the one hand and hardware improvements on the other hand
allow for an increasingly detailed simulation of complex systems.  That
the underlying models are hardly perfect and the model parameters are
barely known with certainty prompts scientist and engineers to conduct
an end-to-end analysis of the encountered errors.  This process, for
which one commonly relies on probability theory, is known as
\emph{uncertainty quantification} (UQ).

Nowadays, the study and possible reduction of uncertainties are
important to virtually all engineering fields.  Some UQ applications can
be found in \cite{Uncertainty:Bijl2013,Uncertainty:Sarkar2017} for
instance.  Due to the prevalence of uncertainties in the mathematical
modeling and numerical simulation of environmental and water systems, UQ
is of special importance to hydrological disciplines
\citep{Physics:Beven2009,Physics:Beven2012}.  Recent reviews of UQ in the
latter context are for example found in
\cite{Hydro:Refsgaard2007,Hydro:Deletic2012,Hydro:DeChant2014}.

The overall goal of UQ is the fair assessment and subsequent improvement
of the accuracy of predictive models.  This provides the basis for
model-assisted decision making
\citep{Hydro:Ascough2008,Hydro:Uusitalo2015} and real-time control
\citep{Hydro:Garcia2015}, which ultimately supports the effective
management of engineered and environmental systems.
In this sense, the \revision{adequate representation}, thorough analysis and treatment of uncertainty is an
ambitious project whose realization typically involves a variety of sub-tasks.
These standard UQ tasks encompass \revision{expert elicitation \citep{Uncertainty:Ayyub2001,Uncertainty:OHagan2006}},
uncertainty  propagation \citep{Hydro:Lei1994,Hydro:Gabellani2007},
sensitivity analysis \citep{Hydro:Baroni2014,Hydro:Pianosi2016}
and Bayesian \revision{analysis} \citep{Hydro:Huard2006,Hydro:Fernandes2017}.

In \emph{forward propagation}, one starts by assigning a probability
distribution to the uncertain parameters of a predictive model.
Afterwards, the goal is to determine the corresponding distribution of
the model response \citep{Uncertainty:Lee2009,Uncertainty:Arnst2014}.
While the input distribution reflects the degree as to which one does
not know the parameter values precisely, the output distribution
quantifies the lack of confidence that we must have in the model
predictions.  A standard method to perform uncertainty forward
propagation is \emph{Monte Carlo} (MC) simulation
\citep{Uncertainty:Helton2006}.  Here, one draws independent samples from
the input distribution and computes the corresponding model responses.
Thereupon, the statistical moments and a histogram-based representation
of the output distribution can be obtained.  MC sampling is very robust
and rests on mild assumptions, yet it is very expensive and suffers from
a slow convergence rate.

The study of how important the various input parameters are with respect
to the model response is called \emph{sensitivity analysis}
\citep{Uncertainty:Saltelli2004,Uncertainty:Saltelli2008}.  There are
certainly different possibilities of defining and analyzing sensitivity
measures.  In a global variance-based analysis one compares and ranks
the input variables according to their contribution to the total
response variance \citep{Uncertainty:Iooss2015,Uncertainty:Prieur2017}.
This does not only provide valuable insight into the input-output
relationship established by the model, it also allows one to restrict
the attention to the most influential parameters and to allocate the
available resources accordingly.  Though, the practical computation of
global sensitivity indices, e.g.\ in a MC sampling framework, is usually
expensive because it requires a moderate to high number of model
evaluations.

\emph{Bayesian inference} is a probabilistic framework for statistical
inference and uncertainty reduction \citep{Nagel:JAIS2015,Nagel:PEM2016}.
The initial uncertainty of the parameters, before the data have been
realized or at least before they are analyzed, is formulated as the
prior probability distribution.  A conditional statistical model of the
observables given values of the unknown parameters, which for the actual
data gives rise to the likelihood function, is constructed.  Eventually,
the posterior distribution ensues from Bayes' law.  This is a
distribution of the unknown parameters conditioned on the acquired
observations, i.e.\ it represents the new state of knowledge and reduced
uncertainty after the information contained in the data has been
processed.  The mean or the mode of the posterior are often taken as
point estimates of the parameters, whereas the spread of the posterior
can be regarded as a measure of the statistical estimation uncertainty.
Bayesian inference does not only establish a consistent foundation for
parameter estimation and inverse modeling, it can also provide solid
answers to the delicate question of model discrepancy
\citep{Bayesian:Kennedy2001,Bayesian:Brynjarsdottir2014}.  Moreover,
model \revision{evidences} can be judged, which in turn enables model
comparison, selection and averaging
\citep{Bayesian:Beck2004,Bayesian:Park2011}.

The chief task in Bayesian inference is the computation of the posterior
distribution.  Sampling-based approaches such as \emph{Markov chain
  Monte Carlo} (MCMC) \citep{MCMC:Robert2004,MCMC:Rubinstein2017} are by
far the most widespread techniques for computational Bayesian inference.
Here, the idea is to construct an ergodic Markov chain that exhibits the
posterior as the stationary distribution.  The posterior can be sampled
that way and all relevant conditional expectation values can be
estimated.  Unfortunately, due to the sample autocorrelation, MCMC is
less efficient than standard MC simulation with independent samples.
Therefore, the procedure usually calls for a  large number of
algorithmic iterations and evaluations of the forward model.  This
causes an excessive computational cost which easily exceeds the
available budget.  A plethora of advanced sampling methods exists, e.g.\
gradient-driven updating schemes with auxiliary variables
\citep{MCMC:Duane1987,MCMC:Betancourt2017:Misc} or population-based
samplers resting on a tempering schedule
\citep{MCMC:Angelikopoulos2012,MCMC:Hadjidoukas2015}.  Beyond that, there
are some fundamentally different alternatives to MCMC sampling, e.g.\
based on variational strategies
\citep{Bayesian:Franck2016,Bayesian:Franck2017}, optimal transportation
theory \citep{Mapping:ElMoselhy2012,Transport:Parno2016} or spectral
likelihood expansions \citep{Nagel:JCP2016}.  But despite these efforts,
Bayesian model calibration remains a costly endeavor.  Additional costs
are even occasioned when the relative evidences of various competing
models have to be assessed \citep{Bayesian:Schoniger2014,MCMC:Liu2016}.

Following these remarks, one has to recognize that UQ tasks are
extremely computing-intensive, especially for hydrological applications
where the underlying processes are often studied by reference to
nonlinear and highly parametrized models.  This brings about long
computation times and complicates all rigorous uncertainty analyses that
require many model runs, i.e.\ forward propagation, sensitivity analysis
and statistical inference.  In the worst case, hydrologists would be
precluded from carrying out important activities, e.g.\ the intelligent
operation of urban infrastructure through state-of-the-art model-based
predictive control methods could be completely prohibited thereby.  It
is also conceivable that, because the evidence-grounded inference of
physical parameters is rendered impossible, parameter values have to be
chosen in an ad-hoc fashion.  This could lead to questionable
predictions and results.

If one does not want to resort to oversimplified models or to discard
some system analyses and control strategies, one may consider employing
a fast \emph{metamodel} or \emph{surrogate model}
\citep{Hydro:Ratto2012,Hydro:Castelletti2012}.  This possibility has
received much attention lately.  A surrogate model is constructed so as
to emulate the input-output relation of the original simulator, i.e.\ it
is a function approximation.  The most important behavioral
characteristics should be reproduced and the emulator must be cheap to
evaluate.  Very often, this goal can be achieved with \emph{Gaussian
  process regression} \cite{Kriging:Santner2003,Kriging:Rasmussen2006}
or a \emph{polynomial chaos expansion} (PCE)
\citep{PCE:LeMaitre2010,PCE:Xiu2010}.  The former is an interpolation
routine that is predicated on the conditioning of a Gaussian process
prior for the simulator.  The latter approach bears on a basis
decomposition of the response random variable into orthogonal
polynomials in the input variables.  In a non-intrusive manner, a
representative sample of model inputs and the associated model responses
can be used to train either of the mentioned emulators.

After a metamodel is constructed, it can replace the simulator in any of
the discussed UQ analyses, e.g.\ in MC simulation for uncertainty
forward propagation.  It turns out that UQ can be significantly
accelerated that way.
The reason is that metamodeling allows one to exploit structures and
regularity properties of the forward problem.  In Gaussian process
regression one can incorporate prior knowledge about the functional
relationship through the mean and covariance function of the Gaussian
process.  Physical understanding and expectations regarding the
smoothness can be integrated hereby.  Polynomial expansions rest on the
assumption that the model response can be represented well in the global
basis.  They allow for finding and utilizing compressibility (i.e.\ the
spectrum of the coefficients decays fast) and sparsity (i.e.\ only a
small fraction of the terms contribute significantly).

Of course, in the construction of a sufficiently accurate and fast
emulator one may also face problems.  The \emph{curse of dimensionality}
stands for a number of phenomena that form obstacles to the study of
high-dimensional systems.  In the first instance, this relates to the
high-dimensionality of the input space, which impedes UQ analyses in
general.  For the computation of PCE-based metamodels in particular, as
mentioned above, one can alleviate the curse by seeking sparsity with
the aid of regularized regression
\citep{PCE:Blatman2011,PCE:Jakeman2015,PCE:Peng2016} and its Bayesian
variants \citep{PCE:Sargsyan2014,PCE:Karagiannis2015,PCE:Shao2017}.
Another problem, that very often arises in the simulation of
time-dependent hydrological processes, is related to high-dimensional
model outputs.  In order to understand and eventually approximate the
functional relationship between the uncertain hydrological parameters
and the corresponding predictions at successive time instants, the
latter have to be seen as different output variables.  Since current
metamodeling techniques treat each of these scalar outputs separately,
and long time series may comprise some ten thousand points, this might
become a critical problem in hydrological applications.

Motivated by the preceding discussion, the goal of this paper is the
development of efficient surrogate models in the context of dynamical
rainfall-runoff simulation, and the subsequent facilitation of the
sensitivity analysis and Bayesian calibration of a slow process-based
urban water simulator.  The following regularity-exploiting strategy is
pursued to that end.  First, principal component analysis is used for
reducing the dimensionality of the combined model output, that comprises
the predicted runoffs at various time instants.  This allows us to
exploit the fact that discharge is a rather well-structured phenomenon,
i.e.\ as function of the random inputs, the discharge at adjacent times
is highly correlated.  Second, the low number of principal components
obtained this way is metamodeled through sparse polynomial expansions.
Thereby we utilize the sparsity found in the basis decomposition of the
hydrological model.  Principal component analysis and polynomial chaos
expansions, each taken by itself, are well-established tools.  Their
synergy potential, however, has attracted little to no attention so far.

Another novelty is that, third, we introduce a new estimator of
time-variant variance-based sensitivity measures in connection with the
chosen emulation approach.  We show that one can obtain the
sensitivities of the original simulator outputs with respect to the
uncertain inputs from the expansion coefficients of the principal
components.  Last, a comprehensive Bayesian analysis of the parametric
uncertainties and modeling errors is conducted.  An explicitly
parametrized term is introduced that represents the systematic model
discrepancy as a function of time.  This term acts additively on the
model predictions and it can be inferred together with the unknown model
parameters.
In order to demonstrate the proposed methods, a case study is executed
that involves a drainage model of an urban catchment area in Switzerland
and real experimental data.  Notwithstanding that all developed methods
are explained and demonstrated on the basis of this specific problem,
they are more generally applicable and can be readily adapted to other
water resources or environmental applications.

The remainder of the paper is organized as follows.  An overview of the
forward problem setup and the information available for the model
calibration is provided in \cref{sec:Hydrology:ProblemSetup}.  The
construction of the multivariate emulator by means of output
dimensionality reduction and sparse polynomial chaos expansions is
described in \cref{sec:Hydrology:SurrogateModeling}.  Details on the
estimation of global variance-based sensitivity measures are given in
\cref{sec:Hydrology:SensitivityAnalysis}.  Bayesian parameter estimation
and predictive model correction are performed in
\cref{sec:Hydrology:BayesianCalibration}.  Finally it is summarized and
concluded in \cref{sec:Hydrology:DiscussionAndConclusion}.

\section{Problem setup} \label{sec:Hydrology:ProblemSetup}
In hydrology one distinguishes between physical process-based and purely
data-driven modeling approaches
\citep{Physics:Singh2002,Physics:Todini2011}.  Due to complicated
interactions of the relevant compartments at several spatial and
temporal scales, the physics-oriented simulation of urban water systems
is a peculiarly uncertainty-prone procedure
\citep{Hydro:Bach2014,Hydro:Salvadore2015}.  This motivates a thorough
error and uncertainty analysis.
In this paper, such an analysis is performed for a computer model which
predicts the time-varying runoff from an urban drainage basin that
receives precipitation, see \cite{Hydro:Machac2015:PhD} for a more
detailed description.  The dynamical simulation of the outflow is based
on a series of rainfall intensity measurements, and it requires
knowledge of model parameters such as the Gauckler--Manning roughness
coefficients or the sub-catchment slopes.  Unfortunately, the values of
those parameters are not precisely known, i.e.\ they are uncertain.
Moreover, even if the input parameters could be known perfectly, the
predicted outputs would be still subject to inevitable model-immanent
errors and discrepancies.

For the analysis and eventual reduction of the aforementioned
uncertainties, we have the following pieces of information at hand.
Bounds and prior distributions of the unknown hydrological parameters
are established.  They represent the parametric uncertainties before the
data are analyzed.  About six hundred measurements of the rainfall
intensity during a precipitation event, observed by a single rain gauge
in the catchment, and the runoff at a single outlet are available.
Besides, we have the precomputed results of roughly two thousand
training runs of the simulator.  They were performed for the recorded
rainfall data, while the uncertain parameters had taken on different
values in each of those runs.  Before we begin with surrogate modeling,
sensitivity analysis and model calibration, this section contains a
brief description of the urban drainage model under consideration and
the experimental data at our disposal.

\begin{figure}[!ht]
  \centering
  \includegraphics[height=\HYDROmapHeight]{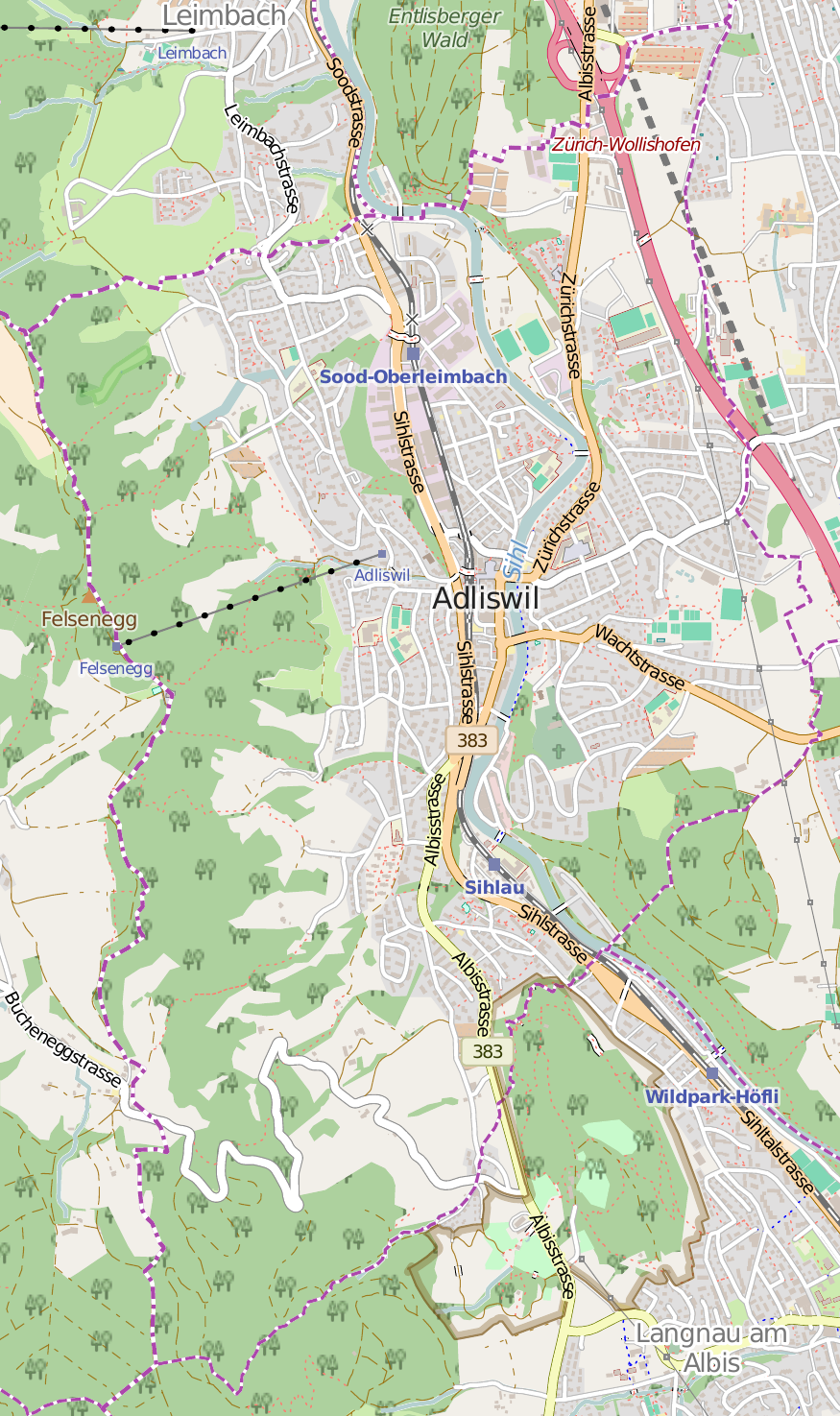}
  \caption{Adliswil, Switzerland (1:50000).
           From \href{https://www.openstreetmap.org}{OpenStreetMap} under \href{https://creativecommons.org/licenses/by-sa/2.0/}{CC BY-SA 2.0}.
           \textcopyright{} \href{https://www.openstreetmap.org/copyright}{OpenStreetMap contributors}.}
  \label{fig:Adliswil}
\end{figure}

The \emph{storm water management model} (SWMM) is a dynamic
rainfall-runoff simulation program for urban areas
\citep{Hydro:SWMM2015}.  It can be used to predict water levels in sewer
manholes and the runoff from catchment areas during dry and wet weather.
A model of the drainage basin of Adliswil, a municipality in the canton
of Z\"{u}rich in the northeast of Switzerland, was developed with the
SWMM at the Swiss Federal Institute of Aquatic Science and Technology.
In \cref{fig:Adliswil} a map is provided that shows the surrounding area
of the size \(\unit[5]{km} \times \unit[3]{km}\).  About
\(\unit[160]{ha}\) of this area, i.e.\ approximately ten percent, are
considered in the model.  As very detailed land-use information is
available, the catchment is represented with approximately one hundred
sub-catchments that are drained by a network consisting of five hundred
pipes.

All sub-catchments and sewer pipes are associated with their own set of
unknown parameters.  This amounts to a fairly large number of unknowns
which is reduced by considering spatial averages only, i.e.\ physical
quantities of the same type are averaged over the sub-catchments or
channels.  The resulting averages thus relate to the catchment as a
whole.  Moreover, the parameters are normalized so as to be
dimensionless and to lie in between reasonable bounds.  A compilation of
the so-obtained scaled parameters \(x_i \in \mathcal{D}_{x_i}\) for \(i
= 1,\ldots,8\) and their bounded domains \(\mathcal{D}_{x_i} =
[\underline{x}_i,\overline{x}_i]\) is presented in
\cref{tab:Parameters}.  Here, the lower and upper bounds are denoted as
\(\underline{x}_i\) and \(\overline{x}_i\), respectively.  The physical
quantities described and their unscaled averages are also provided in
the table.  While the first seven parameters relate to the
sub-catchments, only the last one characterizes the pipes.

\begin{table}[!ht]
  \caption{Hydrological model parameters.}
  \label{tab:Parameters}
  \centering
  \begin{tabular}{llll}
    \toprule
    \(x_i\) & \(\mathcal{D}_{x_i}\) & Physical parameter & Spatial average \\
    \midrule
    \(x_1\) & \([0.5,1.1]\) & Percentage of the impervious area                            & \(\unit[36]{\%}\) \\
    \(x_2\) & \([0.5,1.5]\) & Characteristic width of the overland flow path               & \(\unit[35.7]{m}\) \\
    \(x_3\) & \([0.5,1.5]\) & Slope of the sub-catchments                                  & \(\unit[11.4]{\%}\) \\
    \(x_4\) & \([0.5,1.5]\) & Depression storage height of the impervious area             & \(\unit[2]{mm}\) \\
    \(x_5\) & \([0.5,1.5]\) & Manning roughness coefficient of the impervious area         & \(\unit[0.12]{s \cdot m^{-1/3}}\) \\
    \(x_6\) & \([0.5,1.5]\) & Depression storage height of the pervious area               & \(\unit[2]{mm}\) \\
    \(x_7\) & \([0.5,1.5]\) & Percentage of the impervious area without depression storage & \(\unit[19.04]{\%}\) \\
    \(x_8\) & \([1.0,1.5]\) & Manning roughness coefficient of the channels                & \(\unit[0.012]{s \cdot m^{-1/3}}\) \\
    \bottomrule
  \end{tabular}
\end{table}

\begin{figure}
  \begin{minipage}[b]{.49\linewidth}
    \centering
    \includegraphics[height=\HYDROfigHeight]{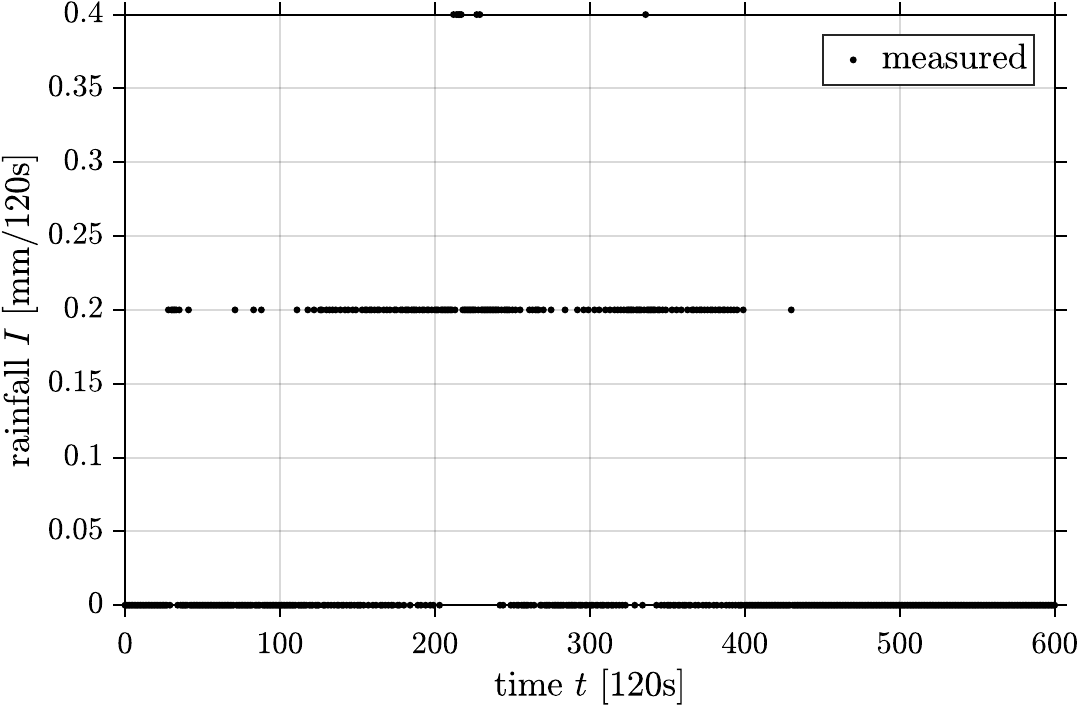}
    \subcaption{Rainfall intensity.}\label{fig:Data:Rainfall}
  \end{minipage}
  \hfill
  \begin{minipage}[b]{.49\linewidth}
    \centering 
    \includegraphics[height=\HYDROfigHeight]{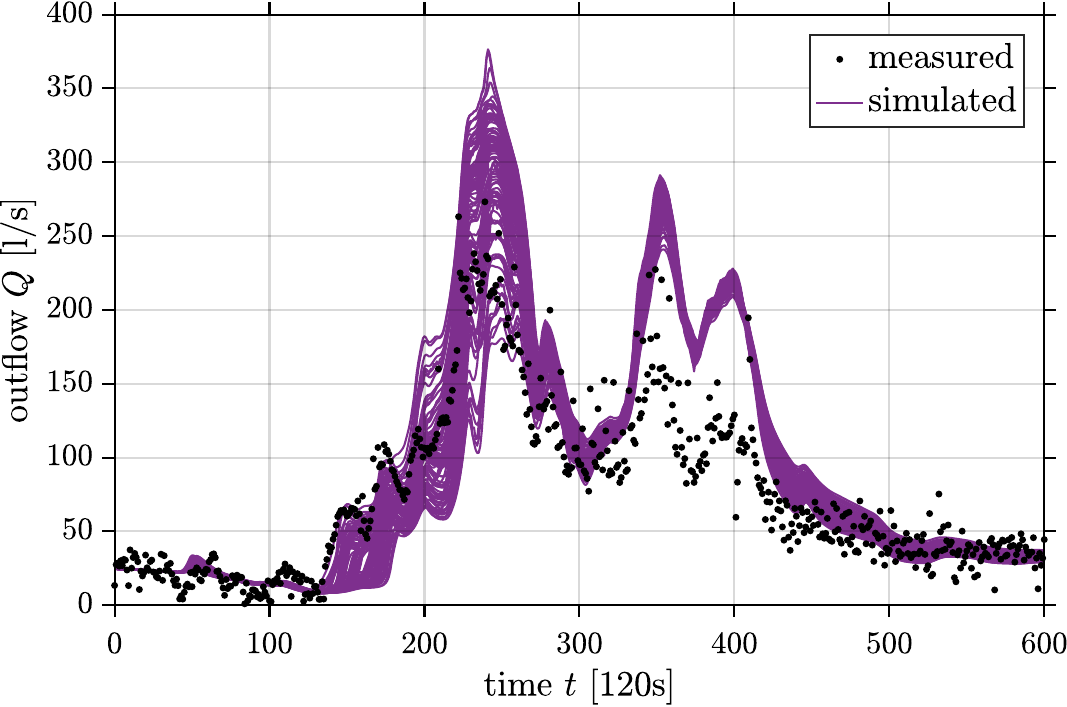}
    \subcaption{Catchment outflow.}
    \label{fig:Data:Outflow}
  \end{minipage}
  \caption{Available data.  Observations of the time-dependent rainfall
    intensity are shown in \cref{fig:Data:Rainfall}.
    Their discrete values reflect the resolution of the measurement
    procedure.  The actually measured outflows during the event can be
    compared against an ensemble of one hundred simulations in
    \cref{fig:Data:Outflow}.  While the simulations
    and measurements are in rough accordance with each other, some
    systematic deviations can be identified in the time interval
    \([150,200]\) and, more obviously, in \([250,500]\).}
  \label{fig:Data} \setcounter{subfigure}{0}
\end{figure}

A single \(15\)-hour rainfall event is considered that had occurred on
May 28, 2013.  Time is denoted as \(t\) in the following.  The
experiment extends over a period with \(t / \unit[120]{s} \in [0,600]\).
Measurements of the varying rainfall intensity \(I\) and the catchment
outflow \(Q\) are taken in regular intervals of two minutes over the
full duration.  For \(i = 0,\ldots,600\) the time instants of the
observations are denoted as \(t_i\).  Both rainfall and outflow
measurements were made at single locations within the drainage basin,
e.g.\ the outflow was measured at the outlet of the system, close to the
wastewater treatment plant.  In Fig.~\ref{fig:Data} the available data are
summarized.  The observations of the rainfall intensity \(I(t_i)\) are
indicated by the black dots in \cref{fig:Data:Rainfall}.  Owing to the resolution of
the measurement device, the recorded intensity takes on discrete values
only.  The recorded outflows \(Q(t_i)\) at the sewage treatment plant
are shown in \cref{fig:Data:Outflow}.

Beyond the observational data just described, the results of
approximately two thousand runs of the SWMM simulator are available.
These will constitute the training runs for the computation of the
surrogate model in the next section.  They were conducted for the
rainfall data shown in \cref{fig:Data:Rainfall} and uniformly
distributed values of the uncertain hydrological parameters.  The
runtime for a single SWMM simulation amounts to approximately twenty
seconds.  For the sake of illustration, a hundred trajectories from
these computer simulations are depicted in \cref{fig:Data:Outflow}.
They can be compared to the actually measured runoffs in the same plot.

The model manages to capture the main trends and characteristics of the
data.  However, in the time interval \(t / \unit[120]{s} \in [150,200]\)
the model systematically underpredicts the outflow.  An even stronger
systematic discrepancy is detected for the time span \(t / \unit[120]{s}
\in [250,500]\) during which the outflow is overpredicted.  It is also
noticed that the model predictions for different values of the uncertain
inputs do not differ significantly, i.e.\ they cannot be discriminated
very well by their ability to trace the data.  The effect is especially
obvious in the second half of the experiment with \(t / \unit[120]{s}
\in [300,600]\).  Here, the mismatch between the data and the model
predictions is apparently dominated by systematic errors and random noise
rather than by variations of the model inputs.
This motivates the study of model discrepancy as a function of time.
\revision{
While random noise can be at least partially attributed to uncertainties in the measurement process,
the systematic errors are a consequence of the model simplifications.
Especially the limitation to spatial parameter averages only,
instead of a more refined representation as spatially variable fields for example,
is believed to be a major factor contributing to the observed time-dependent discrepancies.
}

\section{Surrogate modeling} \label{sec:Hydrology:SurrogateModeling}

In computing-intensive studies it has become common practice to replace
an expensive-to-run simulator of the system by a cheap-to-evaluate
surrogate \citep{Hydro:Ratto2012,Hydro:Castelletti2012}.  These so-called
metamodels or emulators try to mimic the originally defined input-output
relationship as closely as possible.
Dynamic simulators can be emulated in a purely statistics-based manner,
e.g.\ by conditioning a Gaussian process prior on the experimental
design.  Mechanism-based approaches try to enhance the emulation through
an appropriate incorporation of the available physical understanding
\citep{Hydro:Machac2016:b,Hydro:Carbajal2017}.  In particular, the
solution of a simplified problem is incorporated into the prior and
subsequently corrected so as to emulate the full simulator.  An
application of this approach to the Adliswil watershed can be found in
\cite{Hydro:Machac2016:a}.

Polynomial chaos expansions \citep{Hydro:Sochala2013,Hydro:Fan2015}
establish an alternative to Gaussian process--based metamodels.  In this
context, structure-exploiting regression formulations
\citep{PCE:Blatman2011,PCE:Sargsyan2014} do not only capitalize on the
smoothness of the forward model, they also allow one to seek for and
eventually benefit from sparsity in the expansions.  At the same time,
this promises high accuracy and good scalability with the dimensionality
of the input space and the size of the training data.  Examples for the
use of sparse polynomial surrogates in hydrological applications are
found in \cite{PCE:Fajraoui2012,MCMC:Elsheikh2014:b} for instance.

In this section we pursue an integrated strategy that draws on both
dimension reduction and polynomial metamodeling.  First, the
dimensionality of combined model output,
\revision{which consists of all time-indexed predictions,}
is reduced through principal
component analysis.  This allows us to minimize the degree of redundancy
that is inherent in the simulation of a highly structured process such
as the time-variant outflow.  Then, chaos expansions are computed for
the main components as functions of the uncertain inputs, which allows
for revealing and leveraging certain sparsity patterns of the forward
model.  Last, the obtained expansions are combined in order to obtain a
multivariate surrogate model for the full time series of the outflow.
In so doing, the regular properties of the forward problem can be fully
taken advantage of.  That the method is fully non-intrusive, i.e.\ it is
solely based on precomputed input-output pairs, allows for a
straightforward adaptation to other problems.

\subsection{Training runs}
The SWMM implementation of the catchment predicts a complete discharge
time series at the system outlet close to the wastewater treatment plant
throughout the precipitation event.  For the given rainfall intensity
record, the model only acts as a function of the uncertain input
parameters listed in \cref{tab:Parameters}.  Thus we gather the unknown
model input parameters in a vector \(\bm{x} =
(x_1,\ldots,x_\dimParam)^\top\) with \(\dimParam = 8\).  Similarly we
proceed for the rainfall data \(\bm{d} = (d_0,\ldots,d_\dimData)^\top\)
with \(\dimData = 600\).  For \(i = 0,\ldots,\dimData\) we have
introduced \(d_i = I(t_i)\) for the observed rainfall intensities at the
measurement time instants \(t_i\).

The numerical model predicts the vector \(\tilde{\bm{y}} =
(\tilde{y}_0,\ldots,\tilde{y}_\dimData)^\top\) whose entries are the
outflows \(\tilde{y}_i = \tilde{Q}(t_i)\) at the times \(t_i\).  All in
all, \(\tilde{\bm{y}} = \mathcal{M}(\bm{x},\bm{d})\) reflects the
structure of the hydrological simulations.  Since we only consider a
single precipitation event and disregard errors in the rainfall data, we
absorb the dependence on the rainfall into the definition of the forward
model \(\mathcal{M}_{\bm{d}}\) by
\begin{equation} \label{eq:ForwardModel}
  \tilde{\bm{y}} = \mathcal{M}_{\bm{d}}(\bm{x}).
\end{equation}
Note that the dynamical nature of the original system is now completely captured through the pooled multivariate output structure of the model in \cref{eq:ForwardModel}.

We now switch to a probabilistic formulation, where the inputs are
modeled as independent \(\mathcal{D}_{x_i}\)-valued random variables
\(X_i \sim \mathcal{U}(x_i \cond \underline{x}_i,\overline{x}_i)\) with
uniform distributions over the intervals
\([\underline{x}_i,\overline{x}_i]\) for \(i = 1,\ldots,\dimParam\).
The random vector \(\bm{X} \sim \prod_{i=1}^\dimParam \mathcal{U}(x_i
\cond \underline{x}_i,\overline{x}_i)\) with values in
\(\mathcal{D}_{x_1} \times \mathcal{D}_{x_2} \times \ldots \times
\mathcal{D}_{x_\dimParam}\) then represents the input uncertainty.  When
the model \(\mathcal{M}_{\bm{d}}\) is applied to the random inputs
\(\bm{X}\), the output uncertainty is described by the
\(\mathds{R}^{\dimData+1}\)-valued random vector
\begin{equation} \label{eq:OutputRV}
  \tilde{\bm{Y}} = \mathcal{M}_{\bm{d}}(\bm{X}).
\end{equation}
For later considerations, i.e.\ stochastic spectral expansions and variance decompositions,
we will assume from now on tacitly that all components of this random response vector have a finite variance.

In order to construct a metamodel that allows us to cheaply predict to
model response for arbitrary values of the inputs, in total \(K = 2,048\)
training runs were conducted.  Realizations \(\bm{x}^{(k)}\) of the
input variables had to be obtained for \(k = 1,\ldots,K\) and the
corresponding realizations \(\tilde{\bm{y}}^{(k)} =
\mathcal{M}_{\bm{d}}(\bm{x}^{(k)})\) of \cref{eq:OutputRV} had to be
computed.  The inputs were created by Latin hypercube sampling
\citep{MCMC:MacKay1979} in two chunks of \(1,024\) samples each.
Altogether they constitute the \emph{experimental design} \(\mathcal{X}
= (\bm{x}^{(1)},\ldots,\bm{x}^{(K)})\).  The computed model responses,
some of which were already shown in \cref{fig:Data:Outflow}, are
collected into the data matrix
\begin{equation} \label{eq:DataMatrix}
  \mathcal{Y} = \begin{pmatrix}
                  \tilde{\bm{y}}^{{(1)}^\top} \\
                  \tilde{\bm{y}}^{{(2)}^\top} \\
                  \vdots \\
                  \tilde{\bm{y}}^{{(K)}^\top} \\
                \end{pmatrix}
              = \begin{pmatrix}
                  \tilde{y}_0^{(1)} & \tilde{y}_1^{(1)} & \ldots & \tilde{y}_\dimData^{(1)} \\
                  \tilde{y}_0^{(2)} & \tilde{y}_1^{(2)} & \ldots & \tilde{y}_\dimData^{(2)} \\
                  \vdots            & \vdots            & \ddots & \vdots \\
                  \tilde{y}_0^{(K)} & \tilde{y}_1^{(K)} & \ldots & \tilde{y}_\dimData^{(K)} \\
                \end{pmatrix}.
\end{equation}

\subsection{Principal component analysis}
The coordinates of the model output with respect to a certain reference system, e.g.\ the canonical basis, could now be metamodeled individually.
In our case, this would require to handle about six hundred different metamodels simultaneously.
That is inconvenient and may even become infeasible for problems with much longer time series.
Also, it involves a high degree of redundancy, i.e.\ the simulation outputs at contiguous times are highly correlated.

To find a remedy one can choose a basis that is qualified for purposes
of dimension reduction and data compression.  Here we use
\emph{principal component analysis} (PCA) \citep{Statistics:Jolliffe2002}
to that end.  While this technique is mainly used for compressing big
real-world data sets with many features, it can be similarly used for
reducing the model output in the context of computer simulations
\citep{PCE:Sudret2011:Proc,PCE:Sudret2013:Proc}.  A discussion of the
population PCA for a random vector, which is the discrete variant of the
\emph{Karhunen--Lo\`{e}ve} (KL) \emph{expansion} of a stochastic process
\cite{Probability:Loeve1977}, is found at the end of the paper in
Appendix~A.  The empirical sample PCA is recalled in the
following.

Much in the same way as the population PCA works for a random vector,
i.e.\ an orthogonal transformation is applied such that linearly
uncorrelated variables with decreasing variances are obtained, a set of
observed realizations from the random vector is processed in the sample
PCA.  We consider the random vector \(\tilde{\bm{Y}}\) that represents
the model output and a number of realizations \(\mathcal{Y} =
(\tilde{\bm{y}}^{(1)},\ldots,\tilde{\bm{y}}^{(K)})^\top\) from it.
Instead of the exact mean \(\bm{\mu}_{\tilde{\bm{Y}}} =
\mathds{E}[\tilde{\bm{Y}}]\) and covariance matrix
\(\bm{\Sigma}_{\tilde{\bm{Y}}} = \mathrm{Cov}[\tilde{\bm{Y}}]\), which
cannot be determined exactly, one takes their empirical estimates
\begin{equation} \label{eq:EmpiricalMoments}
  \overline{\bm{\mu}}_{\tilde{\bm{Y}}} = \frac{1}{K} \sum\limits_{k=1}^K
  \tilde{\bm{y}}^{(k)}, \quad \overline{\bm{\Sigma}}_{\tilde{\bm{Y}}} =
  \frac{1}{K-1} \sum\limits_{k=1}^K
  (\tilde{\bm{y}}^{(k)}-\overline{\bm{\mu}}_{\tilde{\bm{Y}}})(\tilde{\bm{y}}^{(k)}-\overline{\bm{\mu}}_{\tilde{\bm{Y}}})^\top.
\end{equation}
For \(i = 0,\ldots,\dimData\) the eigenvectors
\(\overline{\bm{\phi}}_i\) and eigenvalues \(\overline{\lambda}_i\) of
the empirical covariance fulfill
\(\overline{\bm{\Sigma}}_{\tilde{\bm{Y}}} \overline{\bm{\phi}}_i =
\overline{\lambda}_i \overline{\bm{\phi}}_i\).  The eigenvalues are
arranged in the descending order \(\overline{\lambda}_0 \geq
\overline{\lambda}_1 \geq \ldots \geq \overline{\lambda}_\dimData\).

Then one finds the smallest \(\dimData^\prime \leq \dimData\) for which
the proportion \(\sum_{i=0}^{\dimData^\prime} \overline{\lambda}_i /
\sum_{i=0}^\dimData \overline{\lambda}_i\) of the total empirical
variance is larger or at least equal than a prespecified threshold.  The
matrix \(\overline{\bm{\Phi}}_{\dimData^\prime} =
(\overline{\bm{\phi}}_0,\overline{\bm{\phi}}_1,\ldots,\overline{\bm{\phi}}_{\dimData^\prime})\)
is composed and for \(k = 1,\ldots,K\) one defines
\begin{equation} \label{eq:CompressedRepresentation}
  \tilde{\bm{z}}^{(k)} = \overline{\bm{\Phi}}_{\dimData^\prime}^\top (\tilde{\bm{y}}^{(k)} - \overline{\bm{\mu}}_{\tilde{\bm{Y}}}).
\end{equation}
This is the reduced PCA representation of \(\tilde{\bm{y}}^{(k)}\) in
terms of the empirical principal components \(\tilde{z}_i^{(k)} =
\overline{\bm{\phi}}_i^\top
(\tilde{\bm{y}}^{(k)}-\overline{\bm{\mu}}_{\tilde{\bm{Y}}})\) for \(i =
0,\ldots,\dimData^\prime\).  The data set is compressed while retaining
most of the total variation by
\begin{equation} \label{eq:CompressedMatrix}
  \mathcal{Z} = \begin{pmatrix}
                  \tilde{\bm{z}}^{{(1)}^\top} \\
                  \tilde{\bm{z}}^{{(2)}^\top} \\
                  \vdots \\
                  \tilde{\bm{z}}^{{(K)}^\top} \\
                \end{pmatrix}
              = \begin{pmatrix}
                  \tilde{z}_0^{(1)} & \tilde{z}_1^{(1)} & \ldots & \tilde{z}_{\dimData^\prime}^{(1)} \\
                  \tilde{z}_0^{(2)} & \tilde{z}_1^{(2)} & \ldots & \tilde{z}_{\dimData^\prime}^{(2)} \\
                  \vdots            & \vdots            & \ddots & \vdots \\
                  \tilde{z}_0^{(K)} & \tilde{z}_1^{(K)} & \ldots & \tilde{z}_{\dimData^\prime}^{(K)} \\
                \end{pmatrix}.
\end{equation}
The compression of the data is lossy, but one can reconstruct the
originally observed samples for \(k = 1,\ldots,K\) approximately as
\begin{equation} \label{eq:ApproximateReconstruction}
  \tilde{\bm{y}}^{(k)} \approx \overline{\bm{\mu}}_{\tilde{\bm{Y}}} +
  \overline{\bm{\Phi}}_{\dimData^\prime} \tilde{\bm{z}}^{(k)} =
  \overline{\bm{\mu}}_{\tilde{\bm{Y}}} +
  \sum\limits_{i=0}^{\dimData^\prime} \tilde{z}_i^{(k)}
  \overline{\bm{\phi}}_i.
\end{equation}

\begin{figure}[!ht]
  \centering
  \includegraphics[height=\HYDROfigHeight]{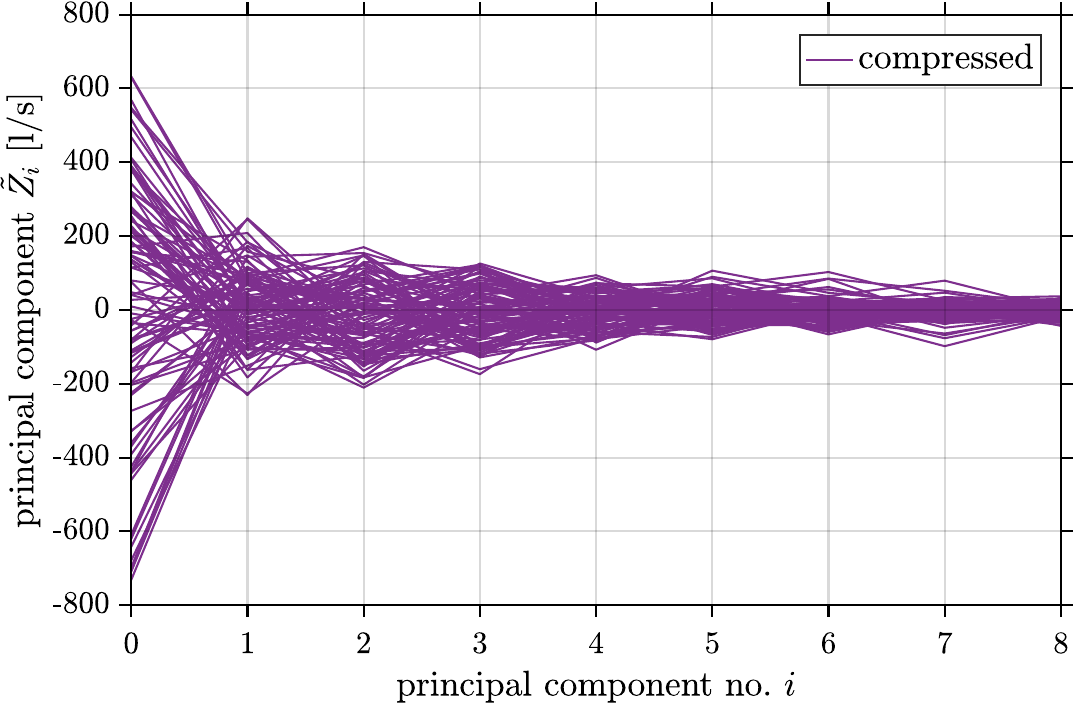}
  \caption{Principal component analysis.  The simulated outflows are
    compressed into the first eight principal components.  This is
    visualized for the hundred realizations that were already depicted
    in \cref{fig:Data:Outflow}.}
  \label{fig:Data:Compression}
\end{figure}

Now we perform PCA to our sample of SWMM simulator responses
\(\mathcal{Y} =
(\tilde{\bm{y}}^{(1)},\ldots,\tilde{\bm{y}}^{(K)})^\top\).  Using
\(\dimData^\prime + 1 = 9\) principal components captures
\(\unit[99]{\%}\) of the total variance of the signal.  Hundred
realizations contained in the compressed data set \(\mathcal{Z} =
(\tilde{\bm{z}}^{(1)},\ldots,\tilde{\bm{z}}^{(K)})^\top\) are visualized
in the parallel coordinate plot in \cref{fig:Data:Compression}.  These
are the empirical principal components of the sample of training runs
that were already shown in \cref{fig:Data:Outflow}.  It can be seen that
the main components are centered around zero and ordered according to
their individual contribution to the total variance.

\subsection{Polynomial chaos expansions}
In order to construct a surrogate model of the \(\dimData + 1\)
components of the forward model response \(\tilde{\bm{Y}}\) as a
function of the unknown parameters \(\bm{X}\), we now only have to
metamodel the first principal components \(\tilde{z}_p(\bm{X}) =
\tilde{Z}_p\) for \(p = 0,\ldots,\dimData^\prime\) in
\cref{eq:TruncatedKLE}.  This can be done through PCEs
\citep{PCE:LeMaitre2010,PCE:Xiu2010}.
Here, one starts with a family of polynomials
\(\{\basis_{\alpha_i}^{(i)}(X_i)\}_{\alpha_i \in \mathds{N}}\) in a
single input variable \(X_i \in \mathcal{D}_{x_i}\) that is indexed by
the polynomial degree \(\alpha_i \in \mathds{N}\).  The polynomials are
assumed to be orthonormal in the sense that
\(\mathds{E}[\basis^{(i)}_{\alpha_i}(X_i) \basis^{(i)}_{\beta_i}(X_i)] =
\delta_{\alpha_i \beta_i}\).

\begin{figure}[htb]
  \centering
  \begin{minipage}[c]{0.45\textwidth}
    \captionof{table}[Orthogonal polynomials]{Orthogonal polynomials.}
    \label{tab:PCE:OrthogonalPolynomials}
    \centering
    \begin{tabular}{lll}
      \toprule
      Distribution & Support & Polynomials \\
      \midrule
      Gaussian & \((-\infty,\infty)\) & Hermite  \\
      Uniform  & \([-1,1]\)           & Legendre \\
      Beta     & \([-1,1]\)           & Jacobi   \\
      Gamma    & \([0, \infty)\)      & Laguerre \\
      \bottomrule
    \end{tabular}
  \end{minipage}%
  \begin{minipage}[c]{0.55\textwidth}
    \centering
    \includegraphics[height=\HYDROfigHeight]{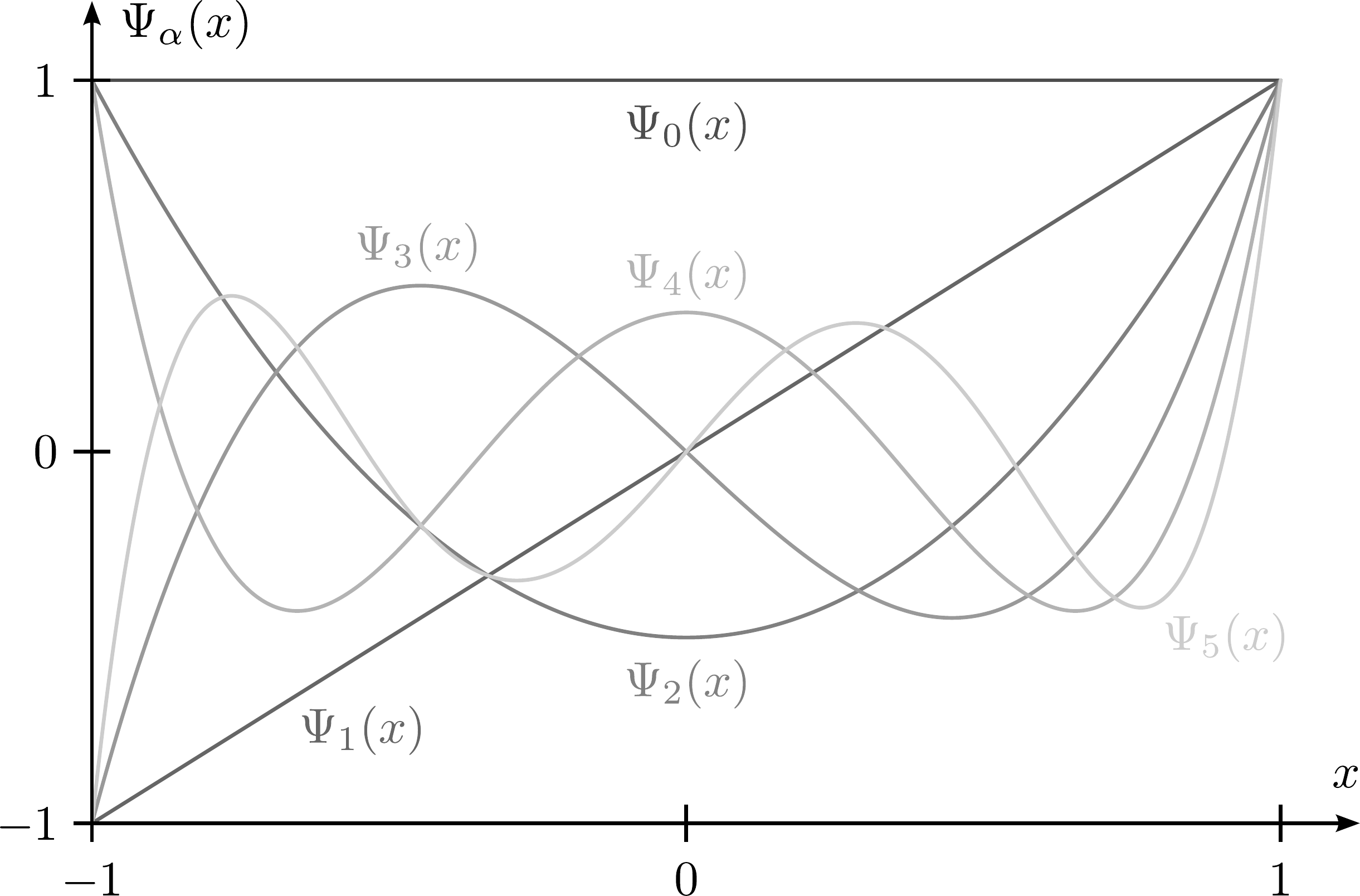}
    \captionof{figure}{Legendre polynomials.}
    \label{fig:PCE:LegendrePolynomials}
  \end{minipage}%
\end{figure}

This way, a couple of well-known probability distributions are
associated with certain families of orthogonal polynomials
\citep{PCE:Xiu2002:b}.  A list of four univariate distributions, their
supports and the corresponding polynomials is provided in
\cref{tab:PCE:OrthogonalPolynomials}.  The first six members of the
Legendre polynomials \(\{\basis_\alpha(x)\}_{\alpha=0}^5\) in a single
variable \(x \in [-1,1]\), that are orthogonal with respect to the
uniform distribution \(\pi(x) = \mathcal{U}(x \cond -1,1)\), are shown
in \cref{fig:PCE:LegendrePolynomials}.  When the random model parameters
do not have a standard form, i.e.\ they do not follow a classical
distribution, one has to transform to \emph{standardized variables} that
follow such a distribution.  As an alternative, one can numerically
construct a sequence of polynomials that are orthogonal with respect to
an arbitrary non-standard input distribution \citep{PCE:Witteveen2007}.

Assuming that the input density factorizes as \(\pi(\bm{x}) = \pi_1(x_1)
\ldots \pi_\dimParam(x_\dimParam)\), one can construct a family
\(\{\basis_{\bm{\alpha}}(\bm{X})\}_{\bm{\alpha} \in
  \mathds{N}^\dimParam}\) of multivariate polynomials by
\(\basis_{\bm{\alpha}}(\bm{X}) = \basis^{(1)}_{\alpha_1}(X_1) \ldots
\basis^{(\dimParam)}_{\alpha_\dimParam}(X_\dimParam)\).  The multi-index
\(\bm{\alpha} = (\alpha_1,\ldots,\alpha_\dimParam) \in
\mathds{N}^\dimParam\) characterizes the polynomials.
This basis can be used to expand the random variables
\(\tilde{z}_p(\bm{X})\) for \(p = 0,\ldots,\dimData^\prime\) with finite
variance as
\begin{equation} \label{eq:PolynomialExpansion}
  \tilde{z}_p(\bm{X}) = \sum\limits_{\bm{\alpha} \in \mathds{N}^\dimParam} \coeffM_{p,\bm{\alpha}} \basis_{\bm{\alpha}}(\bm{X}).
\end{equation}
Here, the set of coefficients \(\{\coeffM_{p,\bm{\alpha}}\}_{\bm{\alpha} \in \mathds{N}^\dimParam}\) is defined through the orthogonal projections
\(\coeffM_{p,\bm{\alpha}} = \mathds{E}[\tilde{z}_p(\bm{X}) \basis_{\bm{\alpha}}(\bm{X})]\).
The series in \cref{eq:PolynomialExpansion} has to be truncated in practice.
This can be done by limiting the total polynomial degree \(\lVert \bm{\alpha} \rVert_1 = \sum_{p=1}^\dimParam \lvert \alpha_p \rvert \leq \maxDegree\) to a certain \(\maxDegree \in \mathds{N}\).
Only terms with \(\bm{\alpha} \in \mathcal{A}_\maxDegree = \{\bm{\beta} \in \mathds{N}^\dimParam \colon \lVert \bm{\beta} \rVert_1 \leq \maxDegree\}\) are then kept in the truncated PCEs
\begin{equation} \label{eq:TruncatedExpansion}
  \tilde{z}_p(\bm{X}) \approx \sum\limits_{\bm{\alpha} \in \mathcal{A}_\maxDegree} \coeffM_{p,\bm{\alpha}} \basis_{\bm{\alpha}}(\bm{X}).
\end{equation}

The truncated expansion in \cref{eq:TruncatedExpansion} is optimal in
the sense that \(\mathds{E}[(\tilde{z}_p(\bm{X}) - \sum_{\bm{\alpha} \in
  \mathcal{A}_\maxDegree} \coeffM_{p,\bm{\alpha}}
\basis_{\bm{\alpha}}(\bm{X}))^2] = \operatorname*{inf}_{\tilde{z}^\star
  \in \mathrm{span}(\{\basis_{\bm{\alpha}}\}_{\bm{\alpha} \in
    \mathcal{A}_\maxDegree})} \mathds{E}[(\tilde{z}_p(\bm{X}) -
\tilde{z}^\star(\bm{X}))^2]\).  Moreover, for \(\maxDegree \rightarrow
\infty\) the series converges in mean square, i.e.\
\(\mathds{E}[(\tilde{z}_p(\bm{X}) - \sum_{\bm{\alpha} \in
  \mathcal{A}_\maxDegree} \coeffM_{p,\bm{\alpha}}
\basis_{\bm{\alpha}}(\bm{X}))^2] \rightarrow 0\).
Another advantage of polynomial chaos--based \revision{representations of black-box models is that
they provide some interpretable insights.}
First of all, one can classify the terms according to
their polynomial degrees and input variables.  Furthermore, their
statistical moments are intimately related to the coefficients of the
expansions.  For example, due to the orthogonality of the polynomial
basis, for the mean and variance one has
\(\mathds{E}[\tilde{z}_p(\bm{X})] = \coeffM_{p,\bm{0}}\) and
\(\mathrm{Var}[\tilde{z}_p(\bm{X})] = \sum_{\bm{\alpha} \in
  \mathds{N}^\dimParam \setminus \{\bm{0}\}}
\coeffM_{p,\bm{\alpha}}^2\), respectively.

For each expansion with \(p = 0,\ldots,\dimData^\prime\), approximations
\(\{\hat{\coeffM}_{p,\bm{\alpha}}\}_{\bm{\alpha} \in
  \mathcal{A}_\maxDegree}\) of the PCE coefficients in
\cref{eq:TruncatedExpansion} can be computed in non-intrusive fashion
\citep{PCE:Xiu2017}.  Pairs of input and output values are processed to
that end.  Here we analyze the representative sample \(\mathcal{X} =
(\bm{x}^{(1)},\ldots,\bm{x}^{(K)})\) of input values, the experimental
design, together with the corresponding values of the principal
components \(\mathcal{Z}_p =
(\tilde{z}_p^{(1)},\ldots,\tilde{z}_p^{(K)})^\top =
(\tilde{z}_p(\bm{x}^{(1)}),\ldots,\tilde{z}_p(\bm{x}^{(K)}))^\top\).
The latter constitute the \(p\)-th column of the matrix \(\mathcal{Z} =
(\mathcal{Z}_0,\ldots,\mathcal{Z}_{\dimData^\prime})\) in
\cref{eq:CompressedMatrix}.  In order to approximate the expansion
coefficients, one can then use linear regression analysis.

We employ \emph{least angle regression} (LAR)
\citep{Statistics:Vidaurre2013,Statistics:Zhang2014}, a powerful
regularized regression technique that promotes sparsity in the PCE
coefficient vectors.  Regressors are penalized in such a way that only
the most dominant ones are retained.  This allows us to mitigate the
curse of dimensionality and has been proven very efficient in the
context of polynomial metamodeling \citep{PCE:Blatman2011}.  We use our
own implementation of the LAR algorithm
\citep{Computing:Marelli2014:Proc,Computing:Uqlab2015:Manual_1104} and
separately compute PCEs of the principal components
\(\tilde{z}_p(\bm{X})\) based on the available experimental design
\(\mathcal{X}\) and the reduced output data \(\mathcal{Z}\).  The
parameters in \cref{tab:Parameters} are linearly transformed so as to
match the uniform standard form in \cref{tab:PCE:OrthogonalPolynomials}.
Following this, normalized multivariate Legendre polynomials in the
standardized random inputs constitute the expansion basis for all PCEs.
As it turns out, the hydrological model is indeed approximately sparse
in the polynomial basis used.  Less than one percent of the total number
of regressors is retained in each of the nine expansions.

The mean squared prediction errors can be used in order to assess the
emulation quality of the sparse PCEs.  However, their empirical
estimation by the average of the squared errors over the experimental
design is overly optimistic.  We therefore use \emph{cross validation}
during the computations, see e.g.\ \cite{PCE:Blatman2010:a} for details.
This protects from overfitting and allows us to adequately estimate how
well the surrogate models generalize beyond the experimental design.
The \emph{leave-one-out errors} for expansions with \(K = 1,024\) \revision{(first batch of samples only)}
and \(K = 2,048\) \revision{(first and second batch combined)}
are reported in \cref{tab:PCE:Errors}.  They are normalized
through division by the sample variance pertaining to the experimental
design.  As expected, the PCE with the richer experimental design
generalizes better than the one with the poorer design for which the
error is approximately twice as high.  One can observe the general trend
that the accuracy of the approximation decays with the order of the
principal components.  Moreover, the metamodeling errors amount to, with
only one exception, less than one percent of the response variance that
is caused by the input uncertainty.  This is deemed accurate enough.

\begin{table}[!ht]
  \caption{Normalized leave-one-out errors.}
  \label{tab:PCE:Errors}
  \centering
  \resizebox{\linewidth}{!}{
  \begin{tabular}{cccccccccc}
    \toprule
    \(K\) & \(\tilde{z}_0\) & \(\tilde{z}_1\) & \(\tilde{z}_2\) & \(\tilde{z}_3\) & \(\tilde{z}_4\) & \(\tilde{z}_5\) & \(\tilde{z}_6\) & \(\tilde{z}_7\) & \(\tilde{z}_8\) \\
    \midrule
    \(1,024\) & \(2.58 \times 10^{-5}\) & \(1.42 \times 10^{-4}\) & \(3.94 \times 10^{-4}\) & \(2.22 \times 10^{-4}\)
             & \(3.37 \times 10^{-3}\) & \(2.70 \times 10^{-3}\) & \(7.35 \times 10^{-3}\) & \(7.37 \times 10^{-3}\) & \(1.11 \times 10^{-2}\) \\
    \(2,048\) & \(1.49 \times 10^{-5}\) & \(6.87 \times 10^{-5}\) & \(1.73 \times 10^{-4}\) & \(1.06 \times 10^{-4}\)
             & \(1.41 \times 10^{-3}\) & \(1.23 \times 10^{-3}\) & \(2.56 \times 10^{-3}\) & \(2.97 \times 10^{-3}\) & \(3.16 \times 10^{-3}\) \\
    \bottomrule
  \end{tabular}
  }
\end{table}

After the computation of a PCE \(\tilde{z}_p(\bm{X}) \approx
\sum_{\bm{\alpha} \in \mathcal{A}_\maxDegree}
\hat{\coeffM}_{p,\bm{\alpha}} \basis_{\bm{\alpha}}(\bm{X})\) for each
principal component with \(p = 0,\ldots,\dimData^\prime\), the random
vector \(\tilde{\bm{Y}} = \mathcal{M}_{\bm{d}}(\bm{X})\) containing the
model outputs can be approximated as
\begin{equation} \label{eq:VectorOutputPCE} \tilde{\bm{Y}} \approx
  \overline{\bm{\mu}}_{\tilde{\bm{Y}}} +
  \sum\limits_{p=0}^{\dimData^\prime} \tilde{z}_p(\bm{X})
  \overline{\bm{\phi}}_p \approx \overline{\bm{\mu}}_{\tilde{\bm{Y}}} +
  \sum\limits_{p=0}^{\dimData^\prime} \left( \sum\limits_{\bm{\alpha}
      \in \mathcal{A}_\maxDegree} \hat{\coeffM}_{p,\bm{\alpha}}
    \basis_{\bm{\alpha}}(\bm{X}) \right) \overline{\bm{\phi}}_p.
\end{equation}
This expansion is henceforth used as a metamodel of the original
response vector.

\begin{figure}[!ht]
  \centering
  \includegraphics[height=\HYDROfigHeight]{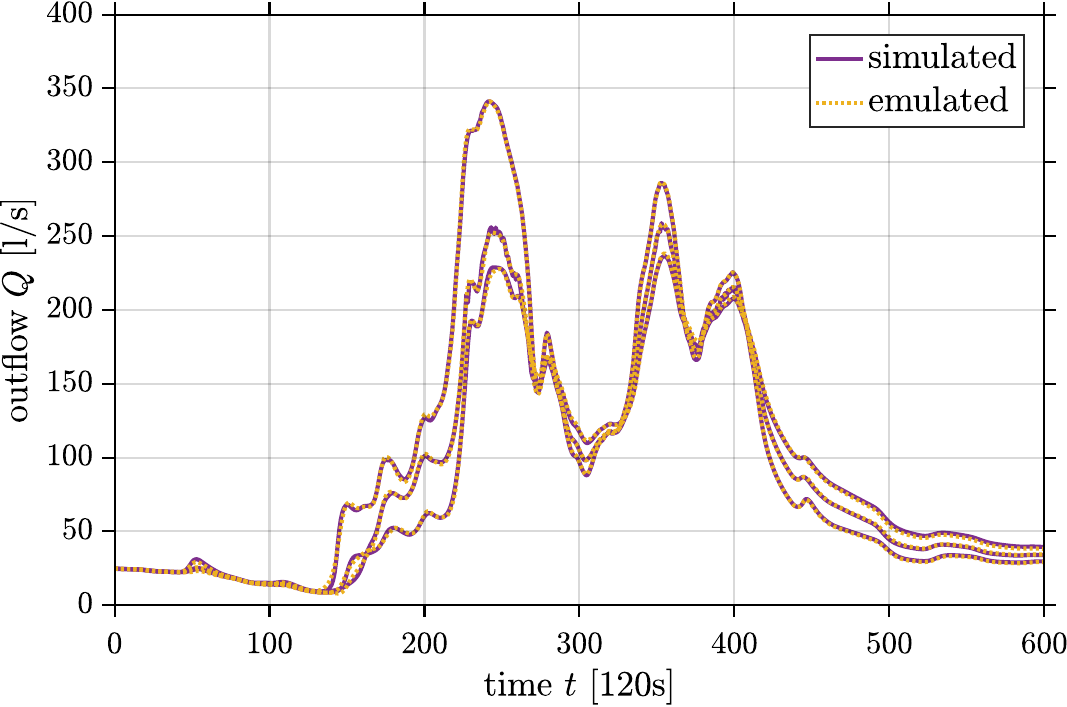}
  \caption{Final metamodel predictions.  Three simulations (solid lines)
    and their approximations (dashed lines) are exemplarily shown.
    Solid and dashed lines that lie on top of each other correspond to
    the same inputs.  Hence, the metamodel is considered as sufficiently
    accurate.}
  \label{fig:PCE:Accuracy}
\end{figure}

In \cref{fig:PCE:Accuracy} the simulated and emulated outflows are shown
for three different input values in the experimental design that were
randomly chosen.  It is again ascertained that the obtained metamodel is
sufficiently accurate for uncertainty quantification purposes.
\revision{
Although our regression approach does not explicitly guarantee the absence of negative outflow predictions,
that would be unphysical, such predictions are not observed in our experiments.
}

\section{Sensitivity analysis} \label{sec:Hydrology:SensitivityAnalysis}

Now we want to perform a global sensitivity analysis of the urban
drainage simulator.  Variance-based sensitivity analysis is based on
apportioning the variance of a model response quantity among several
uncertain input variables
\citep{Uncertainty:Iooss2015,Uncertainty:Prieur2017}.  This provides
relevant insights into the model in question and gives rise to model
reduction schemes \citep{Hydro:Baroni2014,Hydro:Pianosi2016}, e.g.\ one
could construct a simplified model by fixing relatively unimportant
parameters to their nominal values and varying only the most influential
parameters.  For a single response quantity, sensitivity measures such
as Sobol' indices can be efficiently computed with the aid of surrogate
models \citep{Uncertainty:Gratiet2017}, e.g.\ for a PCE one can estimate
the sensitivities by a postprocessing of the expansion coefficients
\citep{PCE:Sudret2008:c}.
In the vector-valued case, one would have to estimate the sensitivity
indices for each output quantity separately.  If the model predictions
are high-dimensional, such as the time series in almost every
hydrological study, this would become inconvenient and time-consuming.

An original solution to this problem is proposed in this section.  It is
based on the combined metamodeling approach presented in the last
section.  Instead of computing the desired Sobol' indices for the
time-variant outflows directly, we consider the indices of the principal
components first.  Then we investigate how the obtained results can be
reused for assessing the sensitivities of the discharge.  As it turns
out, this is feasible through a multivariate extension of the PCE
coefficient postprocessing.  This method for the global sensitivity
analysis of high-dimensional outputs is explained in the following.

\subsection{Sobol' indices}
Recall that we assumed the input variables \(\bm{X} \sim \pi(\bm{x}) =
\pi_1(x_1) \ldots \pi_\dimParam(x_\dimParam)\) to be independent.  For
now, we consider a scalar-valued model response
\(\tilde{Y}_\functionOfInputs = \functionOfInputs(\bm{X})\) with finite
variance, e.g.\ the catchment outflow \(\tilde{Y}_t =
\functionOfInputs(\bm{X})\) at a time instant \(t \in
\{0,\ldots,\dimData\}\) or a principal component \(\tilde{Z}_p =
\functionOfInputs(\bm{X})\) with \(p \in \{0,\ldots,\dimData^\prime\}\).
Variance-based global sensitivity analysis then rests on a decomposition
of the model response into functions with an increasing number of input
parameters.  This so-called \emph{functional analysis of variance}
(ANOVA) or \emph{Hoeffding--Sobol decomposition}
\citep{Uncertainty:Hoeffding1948,Uncertainty:Sobol1993}, which is
sometimes also referred to as a \emph{high-dimensional model
  representation} (HDMR)
\citep{Uncertainty:Rabitz1999:a,Uncertainty:Rabitz1999:b}, can be
written as
\begin{equation} \label{eq:HoeffdingDecomposition}
  \begin{aligned}
    f(\bm{X})
    &= \functionOfInputs_0 + \sum\limits_{1 \leq i \leq \dimParam} \functionOfInputs_i(X_i)
    + \sum\limits_{1 \leq i < j \leq \dimParam} \functionOfInputs_{ij}(X_i,X_j) + \ldots + \functionOfInputs_{1,\ldots,\dimParam}(X_1,\ldots,X_\dimParam) \\
    &= \functionOfInputs_0 + \sum\limits_{1 \leq p \leq \dimParam} \sum\limits_{1 \leq i_1 < \ldots < i_p \leq \dimParam} \functionOfInputs_{i_1,\ldots,i_p}(X_{i_1},\ldots,X_{i_p})
    = \sum\limits_{u \subseteq \{1,\ldots,\dimParam\}} \functionOfInputs_{u}(\bm{X}_u).
  \end{aligned}
\end{equation}
Here, \(\functionOfInputs_0\) is a constant and
\(\functionOfInputs_i(X_i)\) with \(1 \leq i \leq \dimParam\) are
functions of a single variable.  The last-mentioned functions are called
\emph{main effects}.  Functions of more than one variable, such as
\(\functionOfInputs_{ij}(X_i,X_j)\) with \(1 \leq i < j \leq
\dimParam\), are called \emph{interactions}.  An intuitive indexing
scheme for organizing the terms is introduced in the second line of
\cref{eq:HoeffdingDecomposition}.  Here,
\(\functionOfInputs_{u}(\bm{X}_u)\) denotes the interaction of all
variables \(X_{i_q}\) with \(i_q \in u\) that are contained in a certain
set \(u \subseteq \{1,\ldots,\dimParam\}\) and \(f_\varnothing = f_0\)
signifies the constant term.

The Hoeffding--Sobol decomposition is unique if one imposes the vanishing condition that
\(\mathds{E}[\functionOfInputs_u(\bm{X}_u)] = \int \functionOfInputs_u(\bm{x}_u) \, \pi_{i_q}(x_{i_q}) \, \mathrm{d} x_{i_q} = 0\)
for \(\varnothing \neq u \subseteq \{1,\ldots,\dimParam\}\) and \(i_q \in u\).
It follows that all terms but the constant have zero mean and they are mutually uncorrelated, i.e.\ for \(u,v \subseteq \{1,\ldots,\dimParam\}\)
with \(u \neq v\) one has \(\mathds{E}[\functionOfInputs_{u}(\bm{X}_u) \functionOfInputs_{v}(\bm{X}_v)]
= \int \functionOfInputs_{u}(\bm{x}_u) \functionOfInputs_{v}(\bm{x}_v) \, \pi(\bm{x}) \, \mathrm{d} \bm{x} = 0\).
Moreover, one can easily express the unconditional and conditional expectation values of the model response as
\begin{equation} \label{eq:ConditionalExpectations}
  \mathds{E}[\tilde{Y}_\functionOfInputs] = \functionOfInputs_0, \quad
  \mathds{E}[\tilde{Y}_\functionOfInputs \cond X_i] = \functionOfInputs_0 + \functionOfInputs_i(X_i), \quad
  \mathds{E}[\tilde{Y}_\functionOfInputs \cond X_i,X_j] = \functionOfInputs_0 + \functionOfInputs_i(X_i) + \functionOfInputs_j(X_j) + \functionOfInputs_{ij}(X_i,X_j).
\end{equation}
Further conditional expectations are given analogously to \cref{eq:ConditionalExpectations}.
On this basis, one can obtain the terms \(\functionOfInputs_0 = \mathds{E}[\tilde{Y}_\functionOfInputs]\),
\(\functionOfInputs_i(X_i) = \mathds{E}[\tilde{Y}_\functionOfInputs \cond X_i] - \functionOfInputs_0\),
\(\functionOfInputs_{ij}(X_i,X_j) = \mathds{E}[\tilde{Y}_\functionOfInputs \cond X_i,X_j] - \functionOfInputs_0 - \functionOfInputs_i(X_i) - \functionOfInputs_j(X_j)\) and so on recursively.
One often utilizes that a real system model can be well described by an approximation where at most bivariate interactions are included,
i.e.\ \(\functionOfInputs(\bm{X}) \approx \functionOfInputs_0 + \sum_{1 \leq i \leq \dimParam} \functionOfInputs_i(X_i) + \sum_{1 \leq i < j \leq \dimParam} \functionOfInputs_{ij}(X_i,X_j)\).

Note that the ANOVA/HDMR representation contains only finitely many summands, namely \(\sum_{i=0}^\dimParam \binom{\dimParam}{i} = 2^\dimParam\), all of which feature a finite variance.
This allows one to meaningfully partition the variance of the model response \(\mathrm{Var}[\tilde{Y}_\functionOfInputs] = \mathds{E}[\tilde{Y}_\functionOfInputs^2] - \functionOfInputs_0^2\).
The variance \(\mathrm{Var}[\tilde{Y}_\functionOfInputs] = \mathrm{Var}[\sum_{\varnothing \neq u \subseteq \{1,\ldots,\dimParam\}} \functionOfInputs_{u}(\bm{X}_u)]\)
of the sum of the uncorrelated random variables in \cref{eq:HoeffdingDecomposition} is the sum of the individual variances
\begin{equation} \label{eq:VarianceDecomposition}
  \mathrm{Var}[\tilde{Y}_\functionOfInputs]
  = \sum\limits_{1 \leq p \leq \dimParam} \sum\limits_{1 \leq i_1 < \ldots < i_p \leq \dimParam} \mathrm{Var}[\functionOfInputs_{i_1,\ldots,i_p}(X_{i_1},\ldots,X_{i_p})]
  = \sum\limits_{\varnothing \neq u \subseteq \{1,\ldots,\dimParam\}} \mathrm{Var}[\functionOfInputs_{u}(\bm{X}_u)].
\end{equation}
Here, \(\mathrm{Var}[\functionOfInputs_{i_1,\ldots,i_p}(X_{i_1},\ldots,X_{i_p})] = \mathds{E}[\functionOfInputs_{i_1,\ldots,i_p}^2(X_{i_1},\ldots,X_{i_p})]\)
with \(1 \leq p \leq \dimParam\) is termed a \emph{partial variance}.
It quantifies the contribution of the combination of variables \((X_{i_1},\ldots,X_{i_p})\) to the total variance \(\mathrm{Var}[\tilde{Y}_\functionOfInputs]\).
One then defines the \emph{Sobol' index} as the corresponding fraction of the total variance
\begin{equation} \label{eq:SobolIndices}
  S_{i_1,\ldots,i_p} = \frac{\mathrm{Var}[\functionOfInputs_{i_1,\ldots,i_p}(X_{i_1},\ldots,X_{i_p})]}{\mathrm{Var}[\tilde{Y}_\functionOfInputs]}.
\end{equation}

The \emph{first-order indices} \(S_i = \mathrm{Var}[\functionOfInputs_i(X_i)] / \mathrm{Var}[\tilde{Y}_\functionOfInputs]\) with \(1 \leq i \leq \dimParam\) measure the influence of the main effects.
Similarly, the \emph{second-order indices} \(S_{ij} = \mathrm{Var}[\functionOfInputs_{ij}(X_i,X_j)] / \mathrm{Var}[\tilde{Y}_\functionOfInputs]\) with \(1 \leq i < j \leq \dimParam\)
quantify the effect of the bivariate interactions.
In total, there are \(2^\dimParam - 1\) sensitivity indices specified
this way.  They are often summarized by reference to the \emph{total
  Sobol' index} of a variable \(X_{i_q}\) with \(1 \leq i_q \leq
\dimParam\) \citep{Uncertainty:Homma1996}.  It is defined as
\begin{equation} \label{eq:TotalSobolIndices}
  T_{i_q} = \sum\limits_{1 \leq p \leq \dimParam} \sum\limits_{1 \leq i_1 < \ldots < i_q < \ldots < i_p \leq \dimParam} S_{i_1,\ldots,i_q,\ldots,i_p}
  = \sum\limits_{\substack{\varnothing \neq u \subseteq \{1,\ldots,\dimParam\} \\ i_q \in u}} S_u.
\end{equation}
One can interpret the total index as the total effect due to the input parameter \(X_{i_q}\), either in isolation or in conjunction with other variables.
While the indices in \cref{eq:SobolIndices} satisfy \(\sum_{1 \leq p \leq \dimParam} \sum_{1 \leq i_1 < \ldots < i_p \leq \dimParam} S_{i_1,\ldots,i_p} = 1\),
the total Sobol' indices in \cref{eq:TotalSobolIndices} do not have to add up to one.
Instead one has \(0 \leq T_{i_q} \leq 1\) and  \(\sum_{1 \leq i_q \leq \dimParam} T_{i_q} \geq 1\) for the total indices.

Note that the conditional expectations \(\mathds{E}[\tilde{Y}_\functionOfInputs \cond X_i]\) and \(\mathds{E}[\tilde{Y}_\functionOfInputs \cond X_i,X_j]\)
in \cref{eq:ConditionalExpectations} are actually random variables.
This follows from their dependence on \(X_i\) and \(X_j\).
Hence, one can consider the variances of the conditional expectation values.
These are given as
\begin{equation} \label{eq:ConditionalVariances}
  \mathrm{Var}[\mathds{E}[\tilde{Y}_\functionOfInputs \cond X_i]] = \mathrm{Var}[\functionOfInputs_i(X_i)], \quad
  \mathrm{Var}[\mathds{E}[\tilde{Y}_\functionOfInputs \cond X_i,X_j]] = \mathrm{Var}[\functionOfInputs_i(X_i)] + \mathrm{Var}[\functionOfInputs_j(X_j)] + \mathrm{Var}[\functionOfInputs_{ij}(X_i,X_j)].
\end{equation}
While the conditional expectations are related to the terms of the
ANOVA/HDMR decomposition, their variances in
\cref{eq:ConditionalVariances} admit interesting interpretations of the
partial variances and Sobol' indices.  For instance, the first partial
variances are given as \(\mathrm{Var}[\functionOfInputs_i(X_i)] =
\mathrm{Var}[\mathds{E}[\tilde{Y}_\functionOfInputs \cond X_i]]\) and
\(\mathrm{Var}[\functionOfInputs_{ij}(X_i,X_j)] =
\mathrm{Var}[\mathds{E}[\tilde{Y_\functionOfInputs} \cond X_i,X_j]] -
\mathrm{Var}[\functionOfInputs_i(X_i)] -
\mathrm{Var}[\functionOfInputs_j(X_j)]\), respectively.
Consequentially, the Sobol' indices can be expressed as \(S_i =
\mathrm{Var}[\mathds{E}[\tilde{Y}_\functionOfInputs \cond X_i]] /
\mathrm{Var}[\tilde{Y}_\functionOfInputs]\) and \(S_{ij} =
\mathrm{Var}[\mathds{E}[\tilde{Y}_\functionOfInputs \cond X_i,X_j]] /
\mathrm{Var}[\tilde{Y}_\functionOfInputs] - S_i - S_j\).  As a
consequence of \cref{eq:VarianceDecomposition} and the law of total
variance, the total index can be written as \(T_i = 1 -
\mathrm{Var}[\mathds{E}[\tilde{Y}_\functionOfInputs \cond \bm{X}_{\sim
  i}]] / \mathrm{Var}[\tilde{Y}_\functionOfInputs] =
\mathds{E}[\mathrm{Var}[\tilde{Y}_\functionOfInputs \cond \bm{X}_{\sim
  i}]] / \mathrm{Var}[\tilde{Y}_\functionOfInputs]\), where we have
defined \(\bm{X}_{\sim i} =
(X_1,\ldots,X_{i-1},X_{i+1},\ldots,X_\dimParam)\).

\revision{ Notice that for dependent input variables \(\bm{X} \sim
  \pi(\bm{x}) \neq \pi_1(x_1) \ldots \pi_\dimParam(x_\dimParam)\), the
  model representation in \cref{eq:HoeffdingDecomposition} contains
  terms of increasing input dimensionality and, hence, the variance
  decomposition in \cref{eq:VarianceDecomposition}, is not unique
  anymore \citep{Uncertainty:Li2012,Uncertainty:Chastaing2012}.
  Although one could still use \(S_i =
  \mathrm{Var}[\mathds{E}[\tilde{Y}_\functionOfInputs \cond X_i]] /
  \mathrm{Var}[\tilde{Y}_\functionOfInputs]\) as the definition of a
  sensitivity measure, it would reflect both the model structure and the
  input dependencies \citep{Uncertainty:Oakley2004}.  The interpretation
  is therefore not as straightforward as in the case of independent
  inputs.  }

\subsection{Multivariate output}
So far we have discussed the sensitivity analysis of a scalar-valued
response variable \(\tilde{Y}_\functionOfInputs =
\functionOfInputs(\bm{X})\) only, but model responses such as
\(\tilde{\bm{Y}} = \mathcal{M}_{\bm{d}}(\bm{X})\) in \cref{eq:OutputRV}
are often vector-valued.  Two related ways of defining sensitivity
measures for multivariate outputs have been established in the
literature.  First, in
\cite{Uncertainty:Gamboa2013,Uncertainty:Gamboa2014} a multivariate
generalization of the Sobol' indices was proposed that is based on the
trace of a decomposition of the model output covariance matrix.  Second,
in \cite{Uncertainty:Campbell2006} it was suggested to expand the model
outputs in a certain basis first, and then to determine the standard
Sobol' indices of the expansions coefficients with respect to the
uncertain model inputs.  A comparison of those approaches is found in
\cite{Uncertainty:GarciaCabrejo2014}.

Determining the Sobol' indices of the principal components is a special
case of the last-mentioned output decomposition approach
\citep{Uncertainty:Lamboni2009,Uncertainty:Lamboni2011}.  It lends itself
to the multivariate sensitivity analysis of models with time-dependent
\citep{Uncertainty:Sumner2012,Uncertainty:Rohmer2014} or spatially
distributed \citep{Uncertainty:Marrel2011,Uncertainty:Marrel2015}
responses quantities.  However, the sensitivity measures of the
principal components obtained this way, or of artificial coefficients in
general, are difficult to interpret.  We therefore establish a link
between the sensitivities of the expansion coefficients and the ones of
the physically meaningful model outputs.  This possibility has been
often overlooked.

In our case, the principal components \(\tilde{\bm{Z}}\) and the actual model outputs \(\tilde{\bm{Y}}\) are related through the
orthogonal transformation \(\tilde{\bm{Z}} = \bm{\Phi}^\top (\tilde{\bm{Y}} - \bm{\mu}_{\tilde{\bm{Y}}})\) in \cref{eq:LinearTransformation},
and vice versa through \(\tilde{\bm{Y}} = \bm{\mu}_{\tilde{\bm{Y}}} + \bm{\Phi} \tilde{\bm{Z}}\) in \cref{eq:BackTransformation}.
For \(t = 0,\ldots,\dimData\) the model outputs are given as \(\tilde{Y}_t = \mu_{\tilde{Y}_t} + \sum_{p=0}^\dimData \tilde{Z}_p \phi_{p,t}\).
Here, \(\mu_{\tilde{Y}_t}\) is the \(t\)-th entry of the mean vector \(\bm{\mu}_{\tilde{\bm{Y}}}\) and \(\phi_{p,t}\) is the corresponding entry of
the \(p\)-th eigenvector \(\bm{\phi}_p\) of the covariance matrix in \cref{eq:Eigenequation}.
One consequentially has \(\mathds{E}[\tilde{Y}_t \cond X_i] = \mu_{\tilde{Y}_t} + \sum_{p=0}^\dimData \mathds{E}[\tilde{Z}_p \cond X_i] \phi_{p,t}\)
for the conditional expectation of \(\tilde{Y}_t\) with respect to an input \(X_i\) with \(i \in \{1,\ldots,\dimParam\}\).
The variance of this conditional expectation is given as
\begin{equation} \label{eq:RecombinationFirstVariance}
  \mathrm{Var}[\mathds{E}[\tilde{Y}_t \cond X_i]]
  = \sum\limits_{p=0}^\dimData \mathrm{Var}[\mathds{E}[\tilde{Z}_p \cond X_i]] \phi^2_{p,t}
  + 2 \sum\limits_{p < q} \mathrm{Cov}[\mathds{E}[\tilde{Z}_p \cond X_i],\mathds{E}[\tilde{Z}_q \cond X_i]] \phi_{p,t} \phi_{q,t}.
\end{equation}

One can now relate the first-order Sobol' index of the output \(\tilde{Y}_t\) with respect to the input \(X_i\)
to the corresponding indices of the principal components \(\tilde{Z}_p\) for \(p = 0,\ldots,\dimData\).
From \cref{eq:RecombinationFirstVariance} one obtains
\begin{equation} \label{eq:RecombinationFirstSobol}
  S_i^{\tilde{Y}_t} = \frac{\mathrm{Var}[\mathds{E}[\tilde{Y}_t \cond X_i]]}{\mathrm{Var}[\tilde{Y}_t]}
  = \sum\limits_{p=0}^\dimData S_i^{\tilde{Z}_p} \frac{\mathrm{Var}[\tilde{Z}_p]}{\mathrm{Var}[\tilde{Y}_t]} \phi^2_{p,t}
  + 2 \sum\limits_{p < q} \frac{\mathrm{Cov}[\mathds{E}[\tilde{Z}_p \cond X_i],\mathds{E}[\tilde{Z}_q \cond X_i]]}{\mathrm{Var}[\tilde{Y}_t]} \phi_{p,t} \phi_{q,t}.
\end{equation}
Here, \(S_i^{\tilde{Z}_p} = \mathrm{Var}[\mathds{E}[\tilde{Z}_p \cond X_i]] / \mathrm{Var}[\tilde{Z}_p]\)
denotes the Sobol' index of a principal component \(\tilde{Z}_p\) with respect to an input variable \(X_i\).
Hence, \cref{eq:RecombinationFirstSobol} allows one to reuse these indices for the sensitivity analysis of original outputs.
It is noted that the covariances \(\mathrm{Cov}[\mathds{E}[\tilde{Z}_p \cond X_i],\mathds{E}[\tilde{Z}_q \cond X_i]]\) of the conditional expectations
\(\mathds{E}[\tilde{Z}_p \cond X_i]\) and \(\mathds{E}[\tilde{Z}_q \cond X_i]\) for \(p,q = 1,\ldots,\dimData\) with \(p \neq q\) emerge in the above expression.
One can proceed similarly for the second-order and higher indices.
In practice, when instead of \cref{eq:BackTransformation} a reduced number of principal components with \(\dimData^\prime \leq \dimData\) is used in \cref{eq:TruncatedKLE},
one employs the induced approximations of \cref{eq:RecombinationFirstVariance,eq:RecombinationFirstSobol}.

\subsection{Practical computation}
The practical computation of the discussed sensitivity measures in
\cref{eq:SobolIndices,eq:TotalSobolIndices} can be based on Monte Carlo
simulation \citep{Uncertainty:Sobol2001,Uncertainty:Dimov2010}.  As it
was originally proven in \cite{PCE:Sudret2008:c}, however, one
can obtain the Sobol' indices of an output quantity from an appropriate
rearrangement of a PCE of that quantity, too.  This is usually more
efficient and, in our case, it is also more convenient.  Since we employ
a polynomial representation of the principal components, their Sobol'
indices can be estimated right away.  Besides, as it is demonstrated
next, one can extract the corresponding indices of the discharge.  This
allows for a time-variant global sensitivity analysis.

The terms of \(\tilde{z}_p(\bm{X}) = \sum_{\bm{\alpha} \in
  \mathds{N}^\dimParam} \coeffM_{p,\bm{\alpha}}
\basis_{\bm{\alpha}}(\bm{X})\) in \cref{eq:PolynomialExpansion} can be
reordered for all \(p = 0,\dots,\dimData^\prime\) so as to match the
structure of the Hoeffding decomposition in
\cref{eq:HoeffdingDecomposition}, i.e.\ the PCE terms are grouped
according to their input variables.  To that effect, let us define the
set of multi-indices \(\mathcal{A}_u =
\{(\alpha_1,\ldots,\alpha_\dimParam) \in \mathds{N}^\dimParam \colon k
\in u \Leftrightarrow \alpha_k \neq 0\}\) for any non-empty set \(u
\subseteq \{1,\ldots,\dimParam\}\).  Following this, the rearrangement
into terms with an increasing number of input variables is given as
\begin{equation} \label{eq:PCEReordering}
  \tilde{z}_p(\bm{X})
  = \coeffM_{p,\bm{0}} + \sum\limits_{\varnothing \neq u \subseteq \{1,\ldots,\dimParam\}} \sum\limits_{\bm{\alpha} \in \mathcal{A}_u} \coeffM_{p,\bm{\alpha}} \basis_{\bm{\alpha}}(\bm{X})
  = \coeffM_{p,\bm{0}} + \sum\limits_{\varnothing \neq u \subseteq \{1,\ldots,\dimParam\}} \tilde{z}_{p,u}(\bm{X}_u).
\end{equation}
Each summand \(\tilde{z}_{p,u}(\bm{X}) = \sum_{\bm{\alpha} \in
  \mathcal{A}_u} a_{p,\bm{\alpha}} \basis_{\bm{\alpha}}(\bm{X})\) in
\cref{eq:PCEReordering} contains only the PCE terms with multi-indices
\(\bm{\alpha} \in \mathcal{A}_u\).
Since we have \(\mathrm{Var}[\tilde{z}_p(\bm{X})] = \sum_{\bm{\alpha} \in \mathds{N}^\dimParam \setminus \{\bm{0}\}} \coeffM_{p,\bm{\alpha}}^2\) and
\(\mathrm{Var}[\tilde{z}_{p,u}(\bm{X}_u)] = \sum_{\bm{\alpha} \in \mathcal{A}_u} \coeffM_{p,\bm{\alpha}}^2\),
the Sobol' indices in \cref{eq:SobolIndices} can be easily determined from the PCE coefficients.
For instance, the first-order Sobol' index of the principal component \(\tilde{Z}_p\) with respect to the input parameter \(X_i\) is obtained for \(u = \{i\}\) as
\begin{equation} \label{eq:PCESobol}
  S_i^{\tilde{Z}_p} = \frac{\sum_{\bm{\alpha} \in \mathcal{A}_{\{i\}}} a_{p,\bm{\alpha}}^2}{\sum_{\bm{\alpha} \in \mathds{N}^\dimParam \setminus \{\bm{0}\}} \coeffM_{p,\bm{\alpha}}^2}.
\end{equation}

The first-order index \(S_i^{\tilde{Y}_t}\) of the time-variant output \(\tilde{Y}_t\) with respect to an input \(X_i\) is given by \cref{eq:RecombinationFirstSobol}.
This formula contains the covariances \(\mathrm{Cov}[\mathds{E}[\tilde{Z}_p \cond X_i],\mathds{E}[\tilde{Z}_q \cond X_i]]\) of the conditional expectations
\(\mathds{E}[\tilde{Z}_p \cond X_i]\) and \(\mathds{E}[\tilde{Z}_q \cond X_i]\) with \(1 \leq p < q \leq \dimData\).
According to \cref{eq:ConditionalExpectations} one has
\(\mathds{E}[\tilde{Z}_p \cond X_i] = \coeffM_{p,\bm{0}} + \sum_{\bm{\alpha} \in \mathcal{A}_{\{i\}}} \coeffM_{p,\bm{\alpha}} \basis_{\bm{\alpha}}(\bm{X})\)
and an analogous expression for \(\mathds{E}[\tilde{Z}_q \cond X_i]\).
Due to the orthogonality of the polynomial basis, the covariance terms can then be written as
\begin{equation} \label{eq:PCECovarianceTerm}
  \mathrm{Cov}[\mathds{E}[\tilde{Z}_p \cond X_i],\mathds{E}[\tilde{Z}_q \cond X_i]] = \sum\limits_{\bm{\alpha} \in \mathcal{A}_{\{i\}}} \coeffM_{p,\bm{\alpha}} \coeffM_{q,\bm{\alpha}}.
\end{equation}
The first-order index \(S_i^{\tilde{Y}_t}\) is eventually determined by \cref{eq:RecombinationFirstSobol} with \cref{eq:PCESobol,eq:PCECovarianceTerm}.
In practice, instead of the infinite series in \cref{eq:PolynomialExpansion}, one uses the truncated one in \cref{eq:TruncatedExpansion} in order to obtain approximations of the Sobol' indices.

\begin{figure}[!ht]
  \begin{minipage}[b]{.49\linewidth}
    \centering
    \includegraphics[height=\HYDROfigHeight]{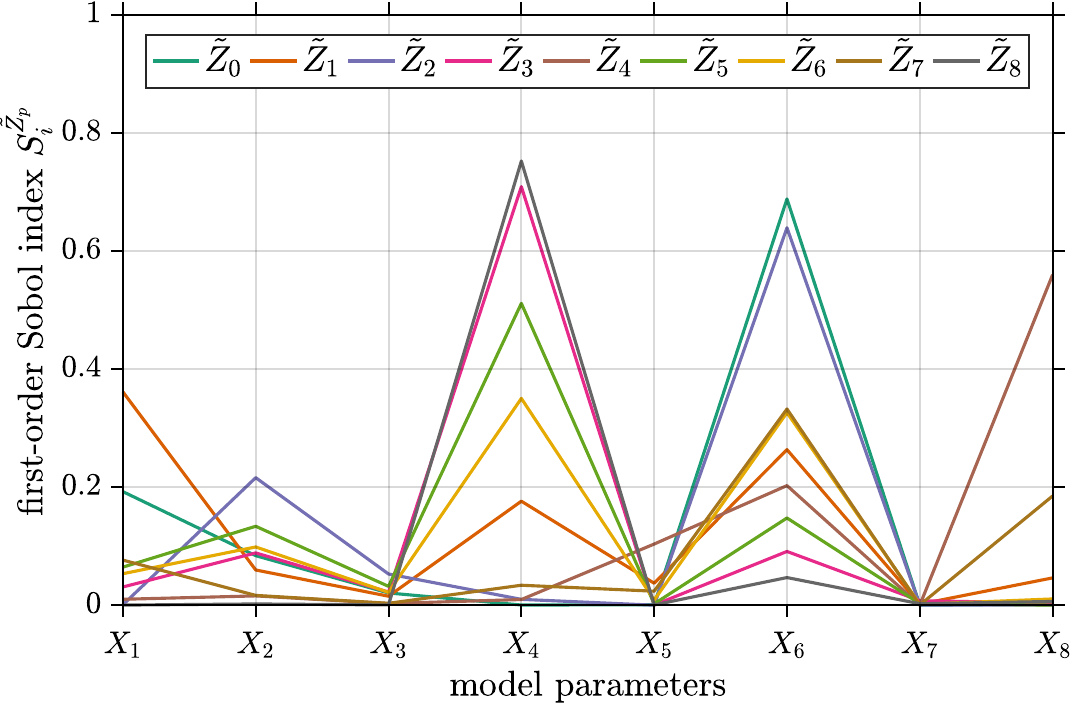}
    \subcaption{Principal components.}     \label{fig:PCE:FirstSobol:Z}
  \end{minipage}
  \hfill
   \begin{minipage}[b]{.49\linewidth}
    \centering
    \includegraphics[height=\HYDROfigHeight]{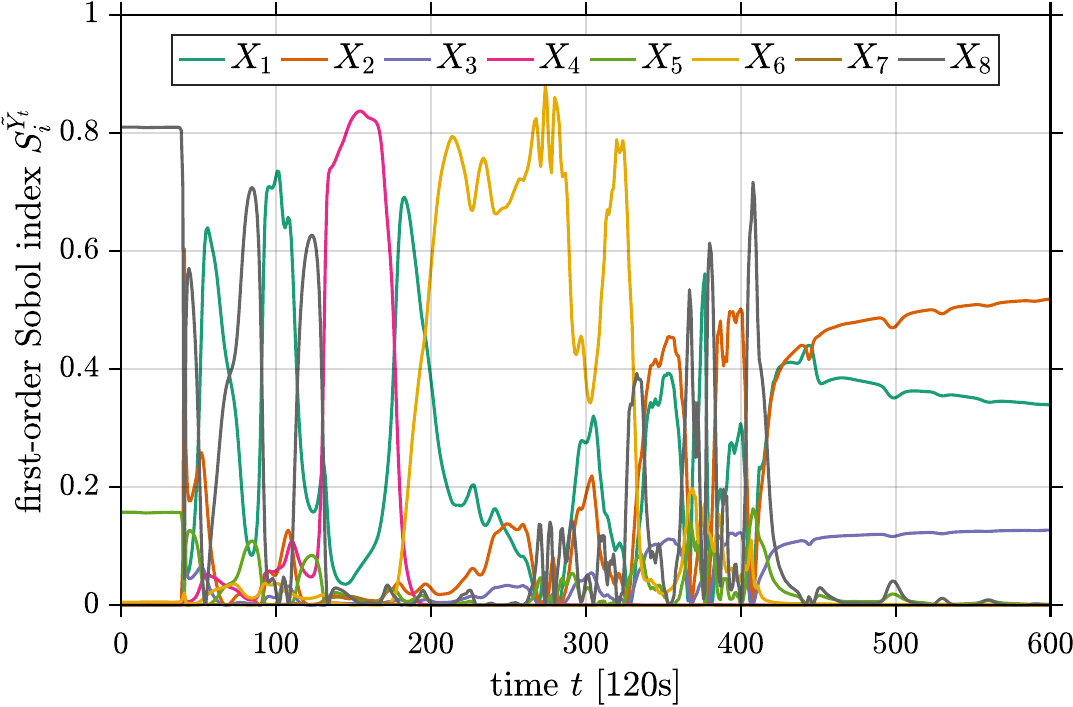}
    \subcaption{Time-variant outflow.}  \label{fig:PCE:FirstSobol:Y}
  \end{minipage}
  \caption{First-order Sobol' indices.  The first Sobol' indices of the
    principal components are shown in \cref{fig:PCE:FirstSobol:Z}.  They
    are obtained by an analysis of the corresponding PCE coefficients.
    The first Sobol' indices of the time-variant outflow are plotted in
    \cref{fig:PCE:FirstSobol:Y} They can be reconstructed from the
    sensitivity indices of the principal components.  While the
    parameters \(X_4\) and \(X_6\) are revealed to be the most
    influential, \(X_7\) contributes insignificantly.}
  \label{fig:PCE:FirstSobol}
\end{figure}

Following this discussion, we compute the Sobol' indices \(S_i^{\tilde{Z}_p}\) and \(S_i^{\tilde{Y}_t}\) for all principal components \(\tilde{Z}_p\) with \(p = 0,\ldots,8\)
and time-dependent discharges \(\tilde{Y}_t\) with \(t = 0,\ldots,600\), with respect to all inputs \(X_i\) with \(i = 1,\ldots,8\).
The results are plotted in \cref{fig:PCE:FirstSobol}.

First of all, \cref{fig:PCE:FirstSobol:Z}
shows the first-order Sobol' indices of the principal components
\(\tilde{Z}_p\) with respect to the input variables \(X_i\).  On the
basis of \cref{eq:PCESobol}, the indices are straightforwardly extracted
from the corresponding PCEs.
Second of all, the first-order Sobol' indices of the time-variant
outputs \(\tilde{Y}_t\) with respect to the inputs are depicted in
\cref{fig:PCE:FirstSobol:Y}.  They are
computed with the aid of
\cref{eq:RecombinationFirstSobol,eq:PCESobol,eq:PCECovarianceTerm} in a
further postprocessing step.

Regarding the contribution to the variances of the principal components,
\cref{fig:PCE:FirstSobol:Z} reveals that the
input parameters \(X_4\) and \(X_6\) are the most important.  They
describe the depression storage height of the impervious and pervious
areas, respectively.  While \(X_6\) influences the first principal
components, \(X_4\) mainly impacts on the last.  The model input
parameter \(X_7\), describing the percentage of the impervious area
without depression storage, hardly contributes at all to the variances
of the principal components.
While the sensitivities of the principal components with respect to the
input parameters are difficult to interpret, the time-evolution of the
Sobol' indices in \cref{fig:PCE:FirstSobol:Y} may be more intuitive to
grasp.  In roughly the interval \(t / \unit[120]{s} \in [100,400]\),
which corresponds to heavy rain and high outflow in \cref{fig:Data}, the
parameters \(X_4\) and \(X_6\) are the most dominant.  In contrast,
\(X_7\) is again the least influential parameter on the whole.

\section{Bayesian calibration} \label{sec:Hydrology:BayesianCalibration}

We now turn towards probabilistic model calibration.  The goal is to
infer the unknown hydrological parameters with the available outflow
data.
Bayesian inference establishes a principled framework for data analysis
and uncertainty reduction \citep{Nagel:JAIS2015,Nagel:PEM2016}.  Unknown
parameters can be inferred from indirectly related measurements.  A
prior probability distribution is assigned to the unknown parameters,
which reflects the epistemic uncertainty of the parameters before the
data are analyzed.  The prior is then conditioned on the realized data,
which gives rise to the posterior distribution.  It represents the
reduced uncertainty after the data have been processed.  Markov chain
Monte Carlo is usually used in order to numerically explore the
posterior \citep{MCMC:Robert2004,MCMC:Rubinstein2017}.
Beyond parametric uncertainties, Bayesian inference allows one to cope
with model discrepancy as well
\citep{Bayesian:Kennedy2001,Bayesian:Brynjarsdottir2014}.  This is
especially important in hydrological applications where predictions are
typically uncertain and biased
\citep{Hydro:DelGiudice2013,Hydro:DelGiudice2015}.

In this section, two different Bayesian models are employed in order to
account for the high level of uncertainty and error in hydrological
model predictions.  The first model is a formulation of nonlinear
inverse modeling with an unknown level of additive noise.  A residual
term following a multivariate Gaussian distribution with zero mean and
unknown variance is used to represent the deviation of the predicted
outflows from the measured values.  The deviations at different times
are assumed to be uncorrelated.  Measurement uncertainty and modeling
errors  are lumped together in this formulation.  The unknown
parameters as well as the unknown error variance can be inferred with
the first simple model.

A more complex Bayesian model is devised that allows us to additionally
capture error correlation and model discrepancy.  The deviations between
the model predictions and the data are now represented as the sum of two
terms.  Similar as in the simple model, the first term represents
Gaussian random errors, but now the errors at different times are
allowed to be correlated.  On top of that, the second term represents a
systematic model discrepancy as a low-order polynomial function of time.
This more realistic model does not only allow us to estimate the
uncertain parameters and the noise level and its correlation structure,
it also enables us to learn the discrepancy function.
\revision{
It is remarked that the discrepancy here pertains to both the model and the analyzed rainfall event,
i.e.\ it captures the systematic error of the model over the duration of the event.
}

\subsection{Independent random errors}
Recall that the measurement data \(\bm{y} = (y_0,\ldots,y_{600})^\top\)
comprise the observations of the outflow \(y_i = Q(t_i)\) for \(i =
0,\ldots,600\).
Assuming that random measurement errors act additively and independently
on the forward model predictions, in the first simple model the measured
data are represented as
\begin{equation} \label{eq:Simple:NoisyData}
  \bm{y} = \mathcal{M}_{\bm{d}}(\bm{x}) + \bm{\varepsilon}.
\end{equation}
Here, \(\bm{\varepsilon}\) is a realization of a random vector with a Gaussian distribution
\(\pi(\bm{\varepsilon} \cond \sigma) = \mathcal{N}(\bm{\varepsilon} \cond \bm{0},\sigma^2 \bm{I})\), where the noise level \(\sigma > 0\) is unknown.
Consequently the following statistical model arises
\begin{equation} \label{eq:Simple:RandomData}
  \pi_1(\bm{y} \cond \bm{x},\sigma) = \mathcal{N}(\bm{y} \cond \mathcal{M}_{\bm{d}}(\bm{x}), \sigma^2 \bm{I}).
\end{equation}
The likelihood function is simply \(\mathcal{L}_1(\bm{x},\sigma) = \mathcal{N}(\bm{y} \cond \mathcal{M}_{\bm{d}}(\bm{x}), \sigma^2 \bm{I})\).
Instead of merely maximizing the likelihood, a fully Bayesian approach is pursued.
For any given prior distribution \(\pi_1(\bm{x},\sigma)\), the corresponding posterior is
\begin{equation} \label{eq:Simple:Posterior}
  \pi_1(\bm{x},\sigma \cond \bm{y}) \propto \mathcal{L}_1(\bm{x},\sigma) \pi_1(\bm{x},\sigma).
\end{equation}

In order to complete the setup, we specify a joint prior of the unknowns
with the product structure \(\pi_1(\bm{x},\sigma) = \pi_1(\bm{x})
\pi_1(\sigma)\) and \(\pi_1(\bm{x}) = \pi_1(x_1) \ldots \pi_1(x_8)\).
\revision{
While the previously used uniform marginals were motivated by coverage considerations,
we now respect the following expert recommendations.
}
The priors for the hydrological
parameters \(x_i \in [\underline{x}_i,\overline{x}_i]\) are normal
distributions \(\pi_1(x_i) = \mathcal{N}(x_i \cond
\mu_{x_i},\sigma_{x_i}^2,\underline{x}_i,\overline{x}_i)\) truncated at
the respective parameter bounds \(\underline{x}_i\) and
\(\overline{x}_i\).  Before the truncation, the distributions are
centered around the midpoint \(\mu_{x_i} =
(\underline{x}_i+\overline{x}_i)/2\) and their standard deviations
\(\sigma_i = (\overline{x}_i-\underline{x}_i)/6\) are set to the sixth
part of the admissible range.  Note that the prior for the hydrological
parameters is different from the uniform distribution that the
experimental design was sampled from.  A uniform distribution
\(\pi_1(\sigma) = \mathcal{U}(\sigma \cond
\underline{\sigma},\overline{\sigma})\) with \(\underline{\sigma} = 0
\times \unit[]{l/s}\) and \(\overline{\sigma} = 100 \times
\unit[]{l/s}\) is selected as the prior for the unknown noise level
\(\sigma\).  The lower bound emerges naturally, whereas the upper bound
is chosen so that it is highly probable that the true or best value is
really contained in the supported interval.

\subsection{Systematic model discrepancy}
The second model is more sophisticated in that it also acknowledges
other sources of uncertainty and error.  In particular, model
discrepancy and random error correlation are captured.  We start the
discussion by representing the measurement data as
\begin{equation} \label{eq:Discrepancy:NoisyData}
  \bm{y} = \mathcal{M}_{\bm{d}}(\bm{x}) + \bm{\delta}(\bm{b}) + \bm{\varepsilon}.
\end{equation}
This is the sum of the model response \(\mathcal{M}_{\bm{d}}(\bm{x})\) at the true \(\bm{x}\) and two other terms that allow for a refined treatment of discrepancy and noise.
The systematic modeling errors are absorbed into the term
\(\bm{\delta}(\bm{b})\), whereas \(\bm{\varepsilon}\) captures the
noise.  We assume that the discrepancy is an unknown function of time
that can be sufficiently well represented as
\begin{equation} \label{eq:Discrepancy:PolynomialDiscrepancy}
  \delta(\bm{b},t) = \sum\limits_{\alpha=0}^\maxDegree b_\alpha \basis_\alpha(t).
\end{equation}
Here, \(\{\basis_\alpha(t)\}_{\alpha=0}^\maxDegree\) is a function basis with \(\numTerms = \maxDegree + 1\) elements
and \(\bm{b} = (b_0,\ldots,b_\maxDegree)^\top\) denotes the unknown coefficients.
The values \(\delta_i(\bm{b}) = \delta(\bm{b},t_i)\) of the discrepancy function at the measurement instants \(t_i\) for \(i = 0,\ldots,600\)
generate the discrepancy vector \(\bm{\delta}(\bm{b}) = (\delta_0(\bm{b}),\ldots,\delta_{600}(\bm{b}))^\top\).

The term \(\bm{\varepsilon}\) is a realization of a random vector
following a multivariate Gaussian distribution \(\pi(\bm{\varepsilon}
\cond \sigma,\tau) = \mathcal{N}(\bm{\varepsilon} \cond
\bm{0},\bm{\Sigma}(\sigma,\tau))\) with an unknown covariance matrix
\(\bm{\Sigma}(\sigma,\tau)\).  For \(i,j = 0,\ldots,600\) the entries of
the covariance matrix are represented as
\begin{equation} \label{eq:Discrepancy:CovarianceMatrix}
  \Sigma_{i,j}(\sigma,\tau) = \sigma^2 \exp \left( - \frac{\lvert t_i - t_j \rvert}{\tau} \right).
\end{equation}
As before, the standard deviation \(\sigma\) determines the noise level.
The additionally introduced correlation length \(\tau\) establishes a
characteristic time scale of the error correlation.  Both parameters
\(\sigma\) and \(\tau\) describing the covariance structure of the error
process are unknown.
In total, we have established the probabilistic data model
\begin{equation} \label{eq:Discrepancy:RandomData}
  \pi_2(\bm{y} \cond \bm{x},\bm{b},\sigma,\tau) = \mathcal{N}(\bm{y} \cond \mathcal{M}_{\bm{d}}(\bm{x}) + \bm{\delta}(\bm{b}), \bm{\Sigma}(\sigma,\tau)).
\end{equation}
The likelihood function \(\mathcal{L}_2(\bm{x},\bm{b},\sigma,\tau) =
\mathcal{N}(\bm{y} \cond \mathcal{M}_{\bm{d}}(\bm{x}) +
\bm{\delta}(\bm{b}), \bm{\Sigma}(\sigma,\tau))\) arises as a result.  If
one has a joint prior \(\pi_2(\bm{x},\bm{b},\sigma,\tau)\), one obtains
the posterior distribution by
\begin{equation} \label{eq:Discrepancy:Posterior}
  \pi_2(\bm{x},\bm{b},\sigma,\tau \cond \bm{y}) \propto \mathcal{L}_2(\bm{x},\bm{b},\sigma,\tau) \pi_2(\bm{x},\bm{b},\sigma,\tau).
\end{equation}

Some prior specifications are now overdue.  We impose a joint prior
distribution with the block-wise independence structure
\(\pi_2(\bm{x},\bm{b},\sigma,\tau) = \pi_2(\bm{x}) \pi_2(\bm{b})
\pi_2(\sigma) \pi_2(\tau)\).  While the priors \(\pi_2(\bm{x}) =
\pi_1(\bm{x})\) and \(\pi_2(\sigma) = \pi_1(\sigma)\) are not altered,
we only have to set \(\pi_2(\bm{b})\) and \(\pi_2(\tau)\).  The latter
is chosen as \(\pi_2(\tau) = \mathcal{U}(\tau \cond
\underline{\tau},\overline{\tau})\) with the lower bound
\(\underline{\tau} = 0 \times \unit[120]{s}\) and a conservatively high
upper bound \(\overline{\tau} = 100 \times \unit[120]{s}\).

We believe that the discrepancy \(\delta(\bm{b},t)\) is a rather smooth
function of time.  It is thus expanded in terms of the first normalized
Legendre polynomials \(\{\basis_\alpha(t)\}_{\alpha=0}^\maxDegree\) up
to rather low degree \(\maxDegree = 5\).  These are the polynomials
shown in \cref{fig:PCE:LegendrePolynomials}.  The time variable is
translated and stretched such that it follows the standard uniform
distribution in \cref{tab:PCE:OrthogonalPolynomials}.  In fact there is
no need to be picky while choosing the polynomial family here.
Since the expansion coefficients \(\bm{b}\) are mere tuning parameters which do not correspond to physically interpretable quantities,
the specification of the prior \(\pi_2(\bm{b}) = \pi_2(b_0) \ldots \pi_2(b_5)\) is a bit delicate.
We opt for Laplace distributions \(\pi_2(b_i) = \mathrm{Laplace}(b_i \cond \mu_{x_i},s_{x_i}) = (2s_{x_i})^{-1} \exp( - \lvert b_i - \mu_{x_i} \rvert / s_{x_i})\) for all \(i = 0,\ldots,5\).
They peak at the mean \(\mu_{x_i} = 0\) and have the scale parameter \(s_{x_i} = 10\) which leads to a standard deviation \(\sigma_{x_i} = s_{x_i} \sqrt{2} \approx 15\).

The double-exponential density \(\mathrm{Laplace}(b_i \cond
\mu_{x_i},s_{x_i})\) decays exponentially with the absolute difference
from the mean, whereas the bell-shaped density \(\mathcal{N}(b_i \cond
\mu_{x_i},\sigma_{x_i}^2) = (2 \pi \sigma_{x_i}^2)^{-1/2} \exp( -
(b_i-\mu_{x_i})^2 / (2\sigma_{x_i}^2) )\) dies down with the squared
difference.  Accordingly, the Laplace distribution has a spikier peak
and fatter tails than the Gaussian at the same time.  Both sparsity of
the coefficient vector and robustness with respect to the prior choice
are promoted that way.  While sparsity is not our main concern at this
point, robustness can be indeed adduced as an argument for the Laplace
prior.  The specification of the scale parameter, however, remains more
or less arbitrary after all.

\subsection{Posterior distributions}
Now we perform fully Bayesian analyses by computing the two posterior
distributions \(\pi_1(\bm{x},\sigma \cond \bm{y})\) and
\(\pi_2(\bm{x},\bm{b},\sigma,\tau \cond \bm{y})\) by means of MCMC
sampling.  The obtained surrogate model \(\hat{\mathcal{M}}_p(\bm{x})\)
is used in place of the original simulator
\(\mathcal{M}_{\bm{d}}(\bm{x})\) throughout the analyses.  A random walk
Metropolis algorithm with a Gaussian proposal distribution is deployed.
Thirty parallel chains with \(10^6\) MCMC iterations are run for both
Bayesian models.  For the first model the parameters \((\bm{x},\sigma)\)
are updated altogether, while for the second model \(\bm{x}\) and
\((\sigma,\tau,\bm{b})\) are updated in two separate blocks.  Roughly
speaking, the posterior computations take half a day for the simple and
about a week for the more complex model.  The non-diagonal covariance
matrix and the block-wise MCMC updates for the second model are
responsible for the runtime difference.

Remember that a single run of the original SWMM simulator takes around
twenty seconds to terminate.  Notwithstanding that this is actually
quite fast, in case the original model would be used in the MCMC
algorithms described above, the total runtime would amount to more than
two hundred days.  On top of the final runs, MCMC always demands the
execution of preliminary runs on the basis of which the algorithm is
tuned and convergence is monitored.  Since this would be obviously
infeasible with the original SWMM simulator, only the employment of the
fast surrogate model here enables us to carry out a Bayesian uncertainty
analysis.

\begin{figure}[p]
  \centering
  \begin{minipage}[b]{.49\linewidth}
    \centering
    \includegraphics[height=\HYDROfigHeightNew]{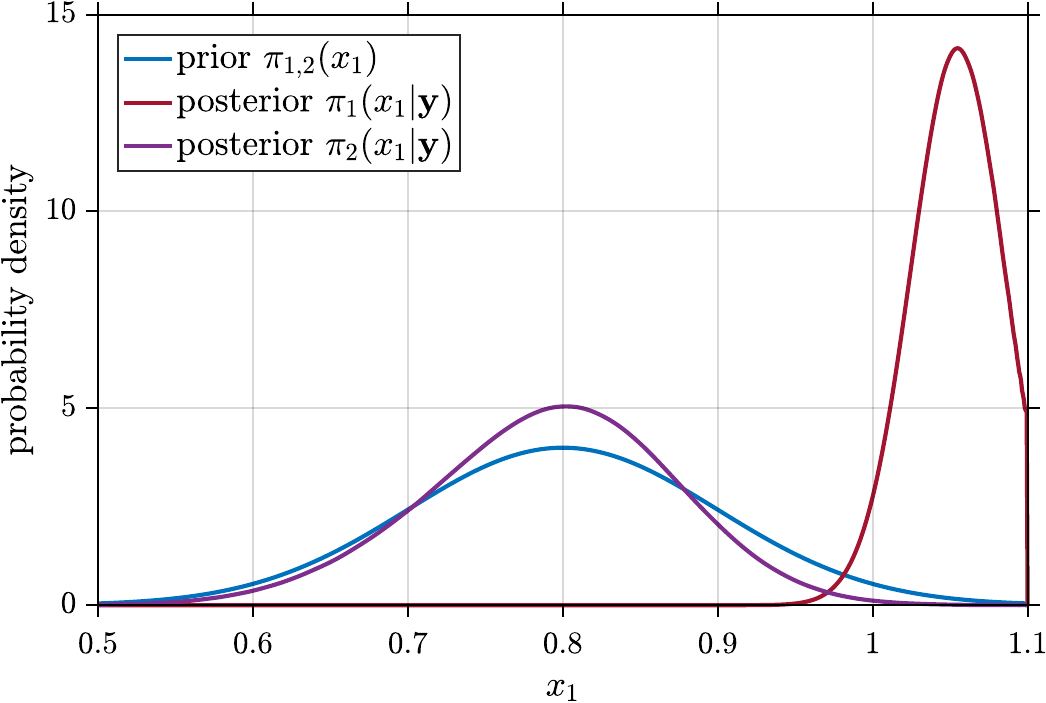}
    \subcaption{Model parameter \(x_1\).}
    \label{fig:Post:x1}
  \end{minipage}\hfill
  \begin{minipage}[b]{\HYDROsubWidth}
    \centering
    \includegraphics[height=\HYDROfigHeightNew]{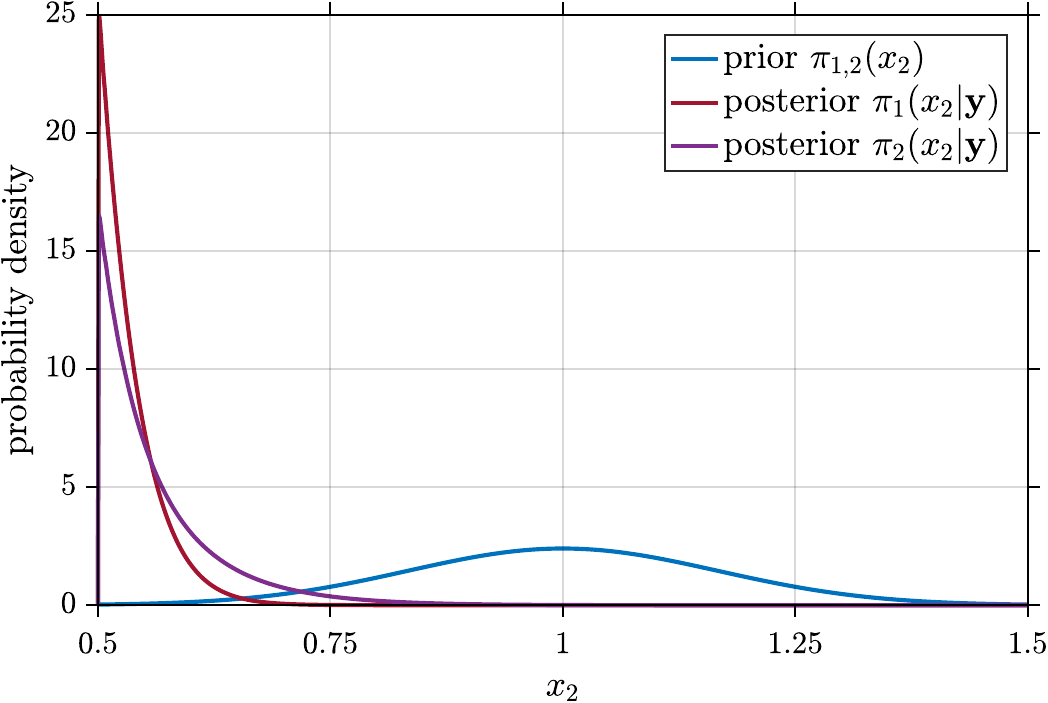}
    \subcaption{Model parameter \(x_2\).}
    \label{fig:Post:x2}
  \end{minipage}\\[3ex]
  \begin{minipage}[b]{\HYDROsubWidth}
    \centering
    \includegraphics[height=\HYDROfigHeightNew]{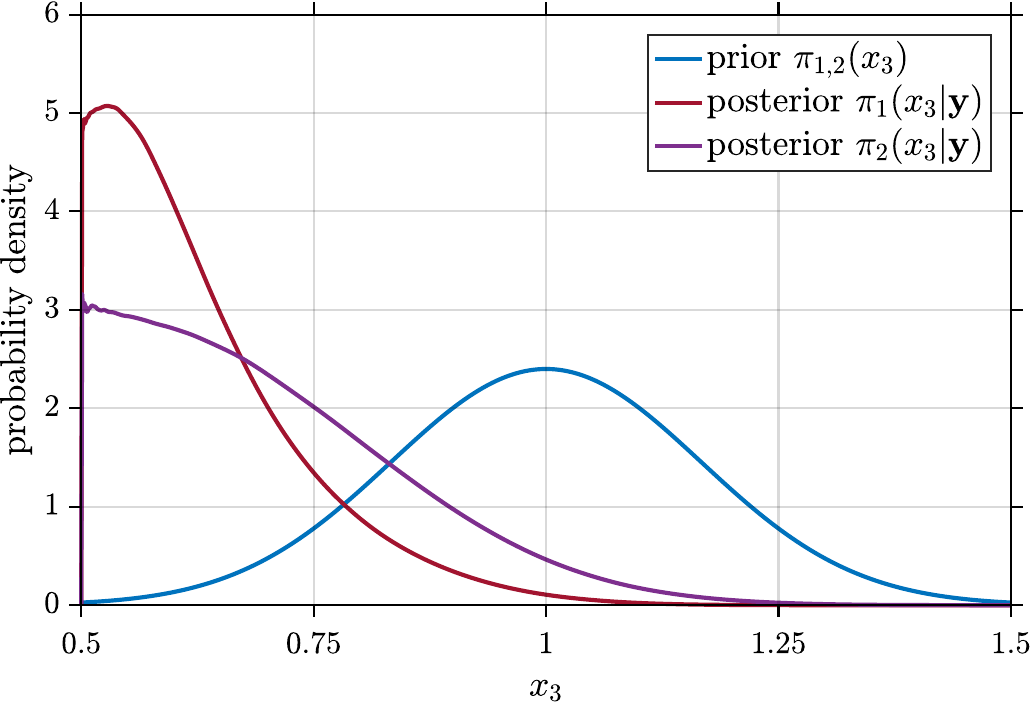}
    \subcaption{Model parameter \(x_3\).}
    \label{fig:Post:x3}
  \end{minipage}\hfill
  \begin{minipage}[b]{\HYDROsubWidth}
    \centering
    \includegraphics[height=\HYDROfigHeightNew]{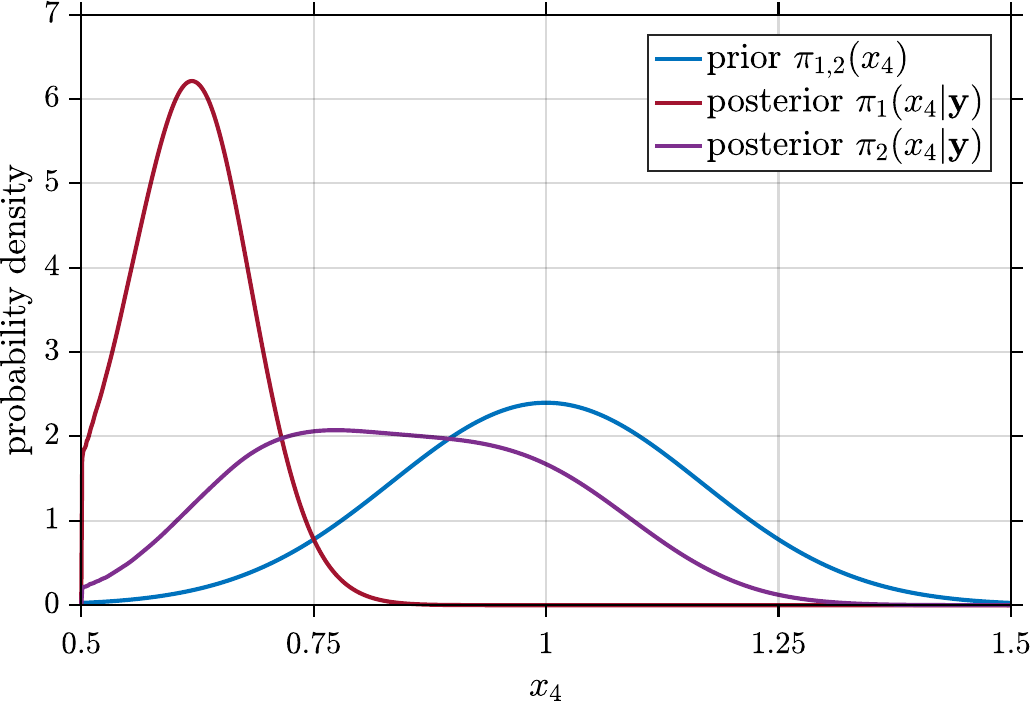}
    \subcaption{Model parameter \(x_4\).}
    \label{fig:Post:x4}
  \end{minipage}\\[3ex]
  \begin{minipage}[b]{\HYDROsubWidth}
    \centering
    \includegraphics[height=\HYDROfigHeightNew]{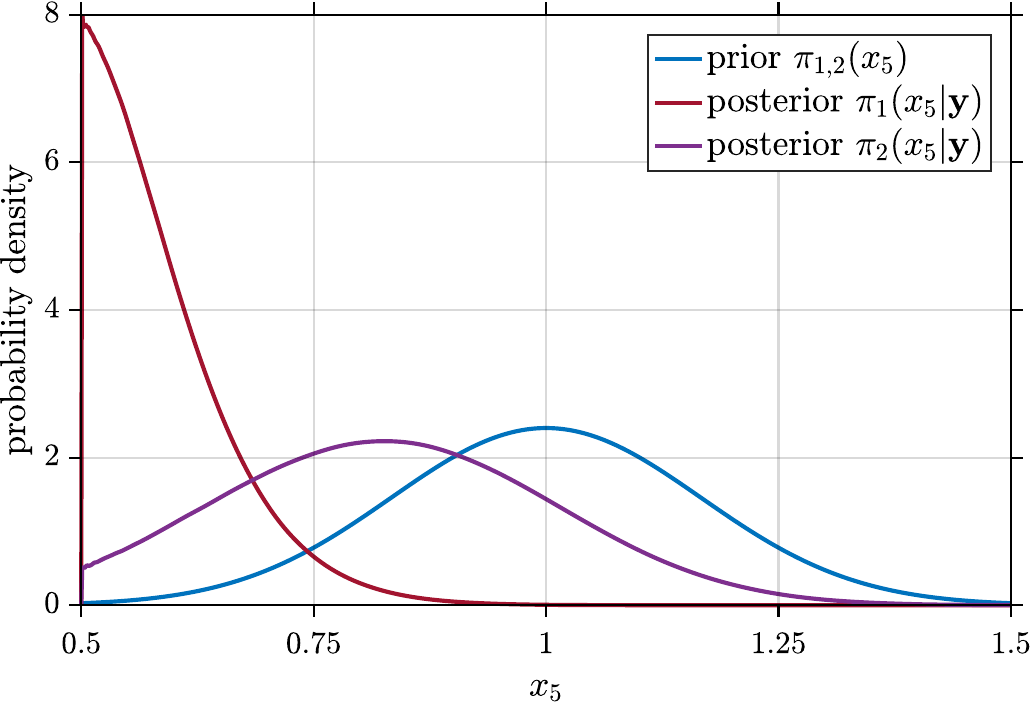}
    \subcaption{Model parameter \(x_5\).}
    \label{fig:Post:x5}
  \end{minipage}\hfill 
  \begin{minipage}[b]{\HYDROsubWidth}
    \centering
    \includegraphics[height=\HYDROfigHeightNew]{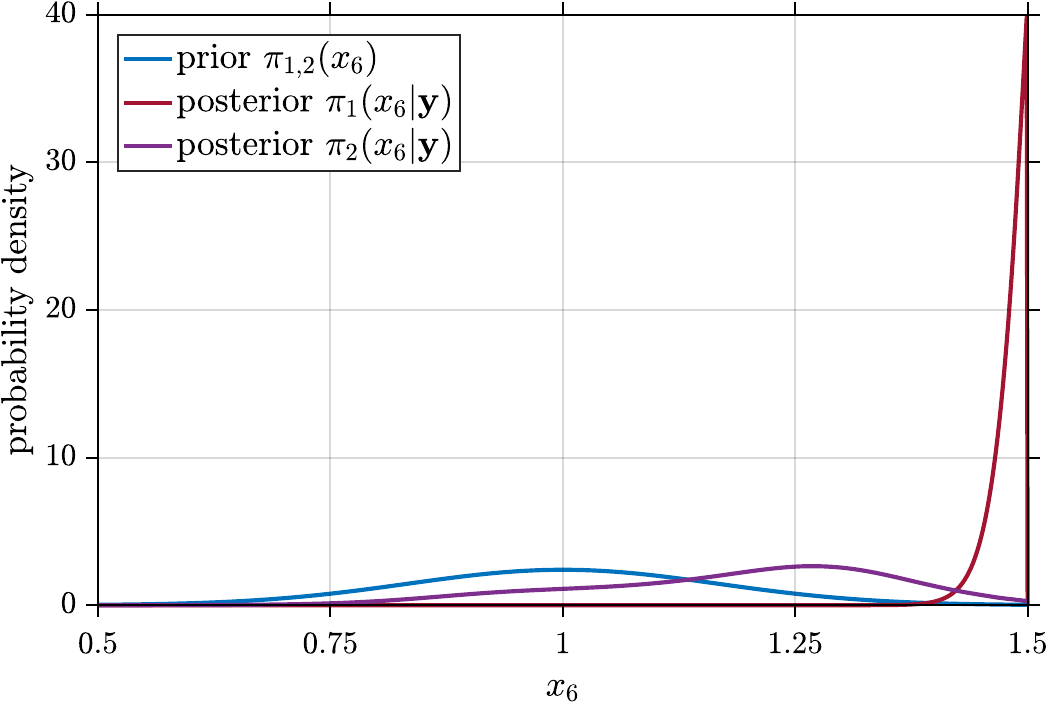}
    \subcaption{Model parameter \(x_6\).}
    \label{fig:Post:x6}
  \end{minipage}
  \begin{minipage}[b]{\HYDROsubWidth}
    \centering
    \includegraphics[height=\HYDROfigHeightNew]{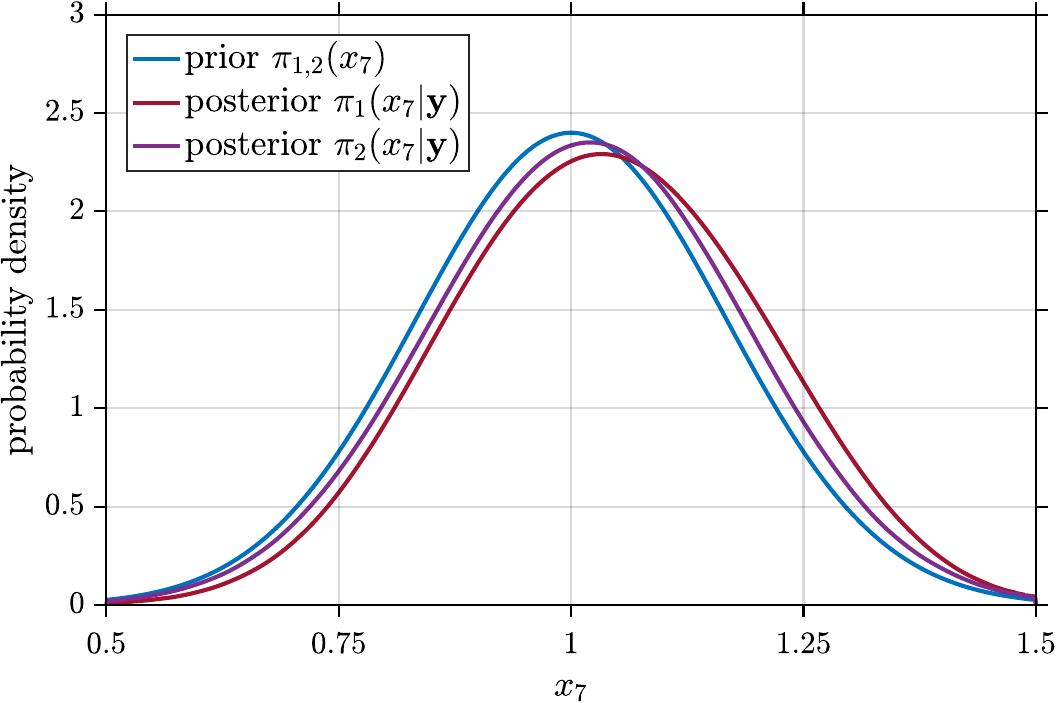}
    \subcaption{Model parameter \(x_7\).}
    \label{fig:Post:x7}
  \end{minipage}\hfill
  \begin{minipage}[b]{\HYDROsubWidth}
    \centering
    \includegraphics[height=\HYDROfigHeightNew]{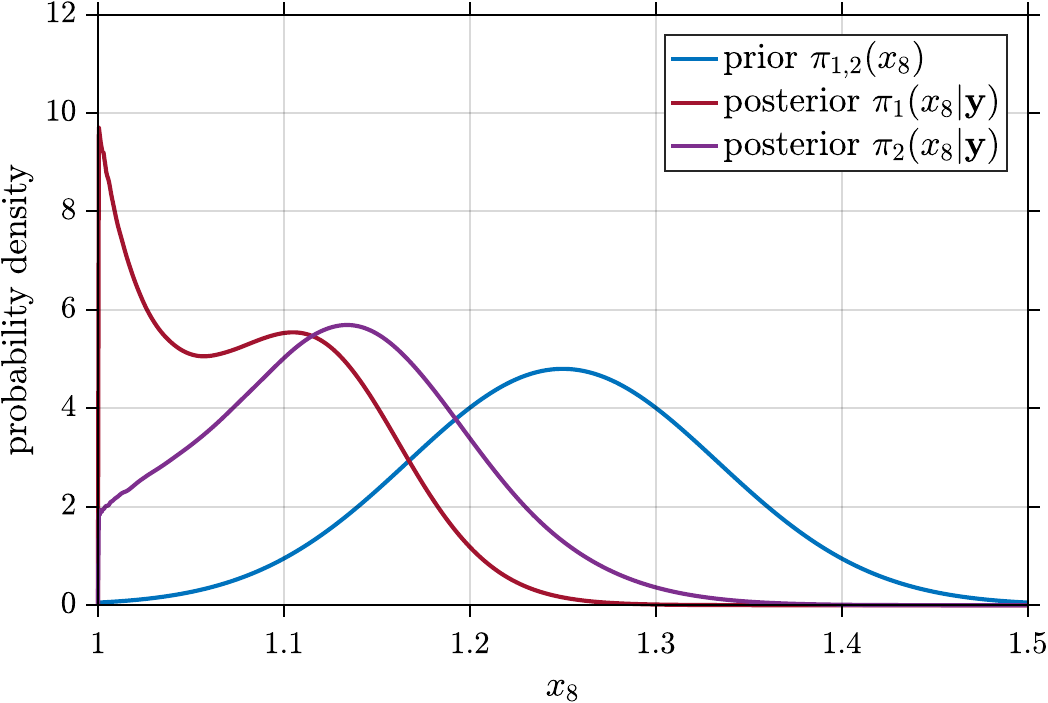}
    \subcaption{Model parameter \(x_8\).}
    \label{fig:Post:x8}
  \end{minipage}
  \caption{Posterior marginals for the hydrological model.  The
    posterior marginals of the parameters \(x_1,\ldots,x_8\) are shown
    above.  While the marginal of \(x_8\) shown in \cref{fig:Post:x8}
    features a distinct posterior structure, the posterior of \(x_7\) in
    \cref{fig:Post:x7} is hardly different from its prior.  The other
    marginals peak close to or directly at one of the parameter bounds.}
  \label{fig:Post:Parameters}
\end{figure}

First of all, we discuss the posterior marginals of the uncertain
hydrological parameters \(\bm{x}\).  All marginals shown in the
following are obtained through kernel density estimation.  For \(i =
1,\ldots,8\) the marginals \(\pi_1(x_i \cond \bm{y})\) and \(\pi_2(x_i
\cond \bm{y})\) of both Bayesian models are plotted in
\cref{fig:Post:Parameters}.
Some marginals of the simple model feature posterior modes close to
their bounds, i.e.\ see \cref{fig:Post:x1,fig:Post:x3,fig:Post:x4} where
the posteriors of \(x_1\), \(x_3\) and \(x_4\) are depicted.  Other
marginals peak directly at the parameter bounds, i.e.\ the marginals of
\(x_2\), \(x_5\) and \(x_6\) in
\cref{fig:Post:x2,fig:Post:x5,fig:Post:x6}, respectively.  The posterior
of \(x_7\) in their bounds, i.e.\ see
\cref{fig:Post:x7} is hardly different from
the prior.  A more complex structure is found in the marginal of \(x_8\)
that is shown in their bounds, i.e.\ see
\cref{fig:Post:x8}.  It has two modes, one of
which peaks at the lower parameter bound.
As compared to the simple model, the marginal posteriors of the second
model with the discrepancy term are generally flattened out and shifted
towards the prior means.

\begin{figure}[!ht]
  \centering
  \begin{minipage}[b]{\HYDROsubWidth}
    \centering
    \includegraphics[height=\HYDROfigHeight]{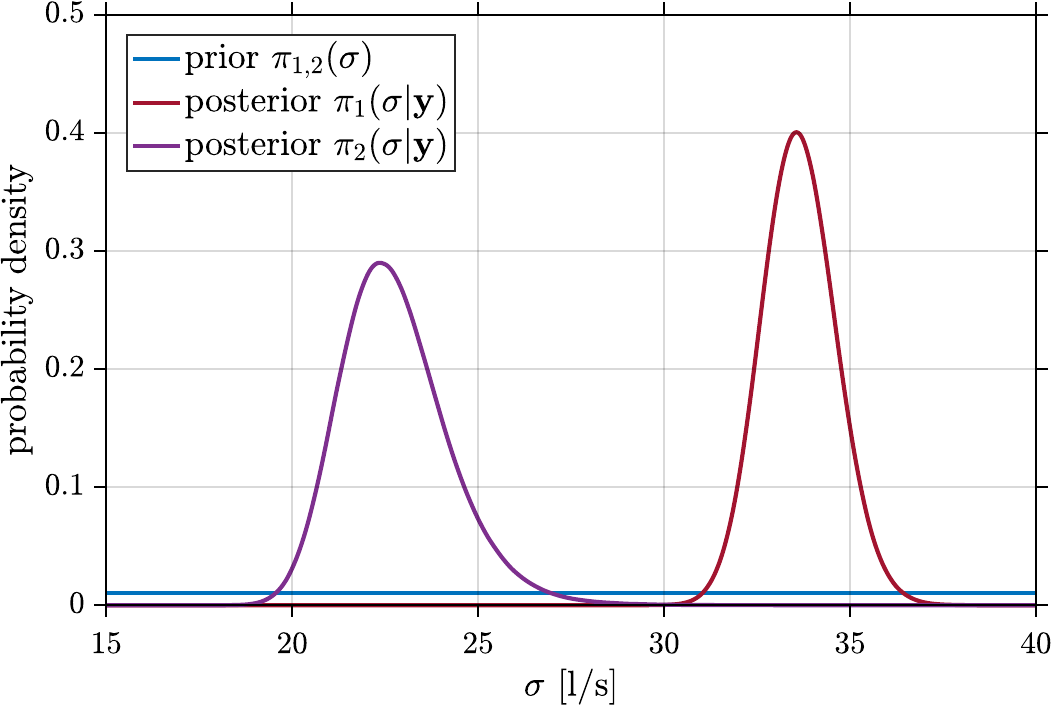}
    \subcaption{Noise level \(\sigma\).}
    \label{fig:Post:Sigma}
  \end{minipage}\hfill
  \begin{minipage}[b]{\HYDROsubWidth}
    \centering
    \includegraphics[height=\HYDROfigHeight]{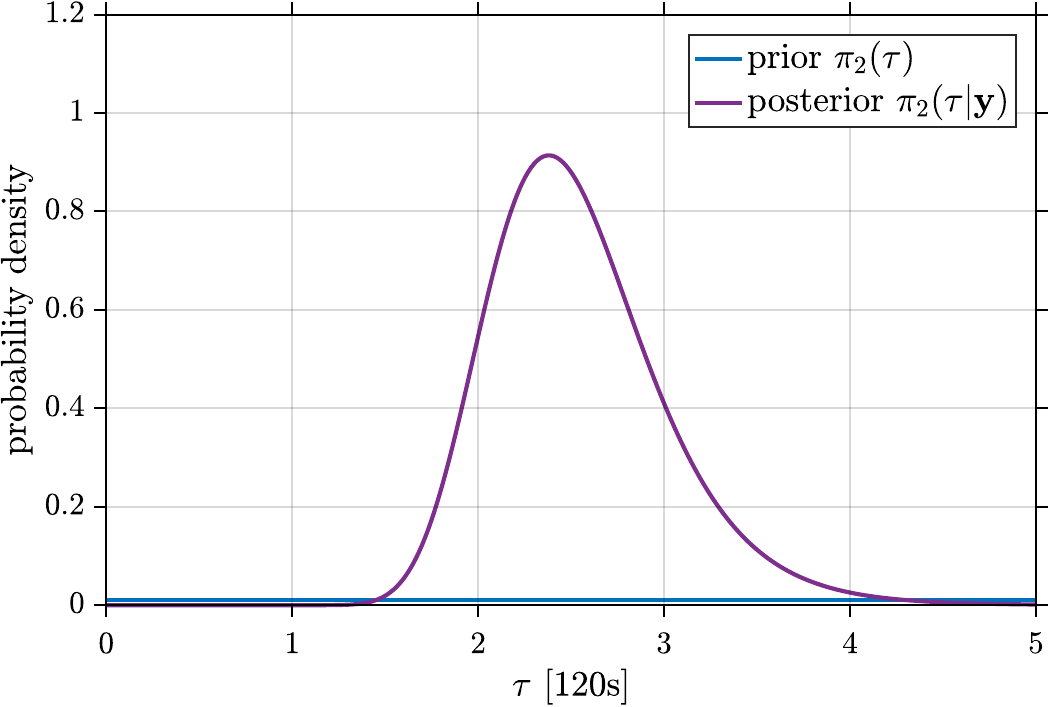}
    \subcaption{Correlation length \(\tau\).}
    \label{fig:Post:Tau}
  \end{minipage}\\[3ex]%
  \begin{minipage}[b]{\HYDROsubWidth}
    \centering
    \includegraphics[height=\HYDROfigHeight]{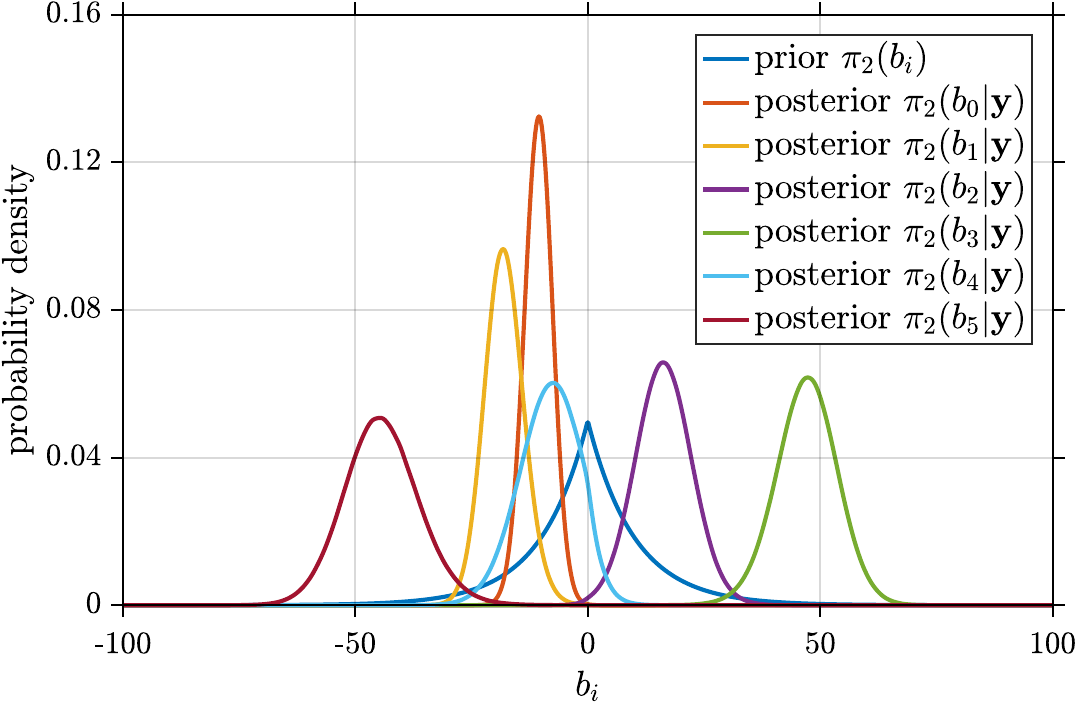}
    \subcaption{Discrepancy coefficients \(b_i\).}
    \label{fig:Post:DiscrepancyModel}
  \end{minipage}%
  \caption{Posterior marginals for the error model.  The posterior
    marginals of the error model parameters \(\sigma, \tau\) and
    \(b_0,\ldots,b_5\) are depicted.  It can be understood from
    \cref{fig:Post:Sigma} that the posterior
    estimate of the noise level parameter \(\sigma\) is higher for the
    first error model than for the second one.  The reason is that the
    latter involves further sources of uncertainty that are parametrized
    through \(\tau\) and \(b_0,\ldots,b_5\).}
  \label{fig:Post:ErrorModel} \setcounter{subfigure}{0}
\end{figure}

The posterior marginals of the parameters describing the random error
model are shown in \cref{fig:Post:ErrorModel}.
As it can be seen from \cref{fig:Post:Sigma}, the marginal
\(\pi_1(\sigma \cond \bm{y})\) suggests a higher value of the standard
deviation \(\sigma\) than the marginal \(\pi_2(\sigma \cond \bm{y})\).
The reason is that according to the first model all errors are
attributed to independent noise only.
In the second model, those errors are also captured by the error
correlation and model discrepancy.  The marginal \(\pi_2(\tau \cond
\bm{y})\) of the correlation length \(\tau\) is plotted in
\cref{fig:Post:Tau}.  It concentrates around a surprisingly low value.
We speculate that the introduction and estimation of the discrepancy
term effectively decorrelates the remaining sources of random error,
which would explain this observation.
In \cref{fig:Post:DiscrepancyModel} all marginals \(\pi_2(b_i \cond
\bm{y})\) of the coefficients \(b_i\) with \(i = 0,\ldots,5\) are shown.
Their actual units are discarded for the sake of simplicity.  It is
interesting to note that the parameters \(\bm{b}\) of the discrepancy
function are estimated quite clearly.  Especially the constant and the
linear term with their coefficients \(b_0\) and \(b_1\) have pronounced
posterior shapes.

Some summaries of the posterior distributions \(\pi_1(\bm{x},\sigma
\cond \bm{y})\) and \(\pi_2(\bm{x},\bm{b},\sigma,\tau \cond \bm{y})\)
are compiled in \cref{tab:Post:Summaries}.  These are point estimates of
the unknown parameters, e.g.\ the posterior mean vectors
\((\hat{\bm{x}},\hat{\sigma}) = \mathds{E}[\bm{x},\sigma \cond \bm{y}]\)
and \((\hat{\bm{x}},\hat{\bm{b}},\hat{\sigma},\hat{\tau}) =
\mathds{E}[\bm{x},\bm{b},\sigma,\tau \cond \bm{y}]\).  Quantities whose
dimension does not equal one are expressed in comparison to the units
that were previously adopted.  Posteriors that peak at the prior bounds
are not summarized well by their mean values only.  Therefore the modes
\((\hat{\bm{x}},\hat{\sigma})_{\mathrm{MAP}} =
\operatorname{arg\,max}_{\bm{x},\sigma} \pi_1(\bm{x},\sigma \cond
\bm{y})\) and
\((\hat{\bm{x}},\hat{\bm{b}},\hat{\sigma},\hat{\tau})_{\mathrm{MAP}} =
\operatorname{arg\,max}_{\bm{x},\bm{b},\sigma,\tau}
\pi_2(\bm{x},\bm{b},\sigma,\tau \cond \bm{y})\) of the joint posterior
densities are shown, too.  They have been obtained through maximizing
the logarithms of the unnormalized posterior densities, i.e.\ the
log--likelihood function plus the log--prior density.  Note that the
individual components of the joint posterior density mode do not have to
coincide with the maxima of the marginal densities.

\begin{table}[!ht]
  \caption{Posterior summaries.}
  \label{tab:Post:Summaries}
  \centering
  \resizebox{\linewidth}{!}{
  \begin{tabular}{llcccccccccccccccc}
    \toprule
    & & \(\hat{x}_1\) & \(\hat{x}_2\) & \(\hat{x}_3\) & \(\hat{x}_4\) & \(\hat{x}_5\) & \(\hat{x}_6\) & \(\hat{x}_7\) & \(\hat{x}_8\)
    & \(\hat{\sigma}\) & \(\hat{\tau}\) & \(\hat{b}_0\) & \(\hat{b}_1\) & \(\hat{b}_2\) & \(\hat{b}_3\) & \(\hat{b}_4\) & \(\hat{b}_5\) \\
    \midrule
    \multirow{2}{*}{\(\pi_1(\cdot \cond \bm{y})\)}
    & Mean
    & \(1.05\) & \(0.54\) & \(0.63\) & \(0.62\) & \(0.59\) & \(1.48\) & \(1.04\) & \(1.09\) & \(33.63\)
    & - & - & - & - & - & - & - \\
    & Mode
    & \(1.06\) & \(0.50\) & \(0.55\) & \(0.62\) & \(0.50\) & \(1.49\) & \(0.99\) & \(1.00\) & \(33.04\)
    & - & - & - & - & - & - & - \\
    \multirow{2}{*}{\(\pi_2(\cdot \cond \bm{y})\)}
    & Mean
    & \(0.79\) & \(0.56\) & \(0.70\) & \(0.85\) & \(0.84\) & \(1.18\) & \(1.02\) & \(1.13\) & \(22.78\) & \(2.53\) & \(-10.63\) & \(-17.97\) & \(16.33\) & \(47.05\) & \(-8.38\) & \(-44.31\) \\
    & Mode
    & \(0.71\) & \(0.50\) & \(0.50\) & \(0.71\) & \(0.59\) & \(0.91\) & \(1.01\) & \(1.03\) & \(19.95\) & \(1.80\) & \(-13.33\) & \(-20.25\) & \(19.39\) & \(46.70\) & \(-10.25\) & \(-42.72\) \\
    \bottomrule
  \end{tabular}
  }
\end{table}

\begin{figure}[!ht]
  \centering
  \begin{minipage}[b]{\HYDROsubWidth}
    \centering
    \includegraphics[height=\HYDROfigHeight]{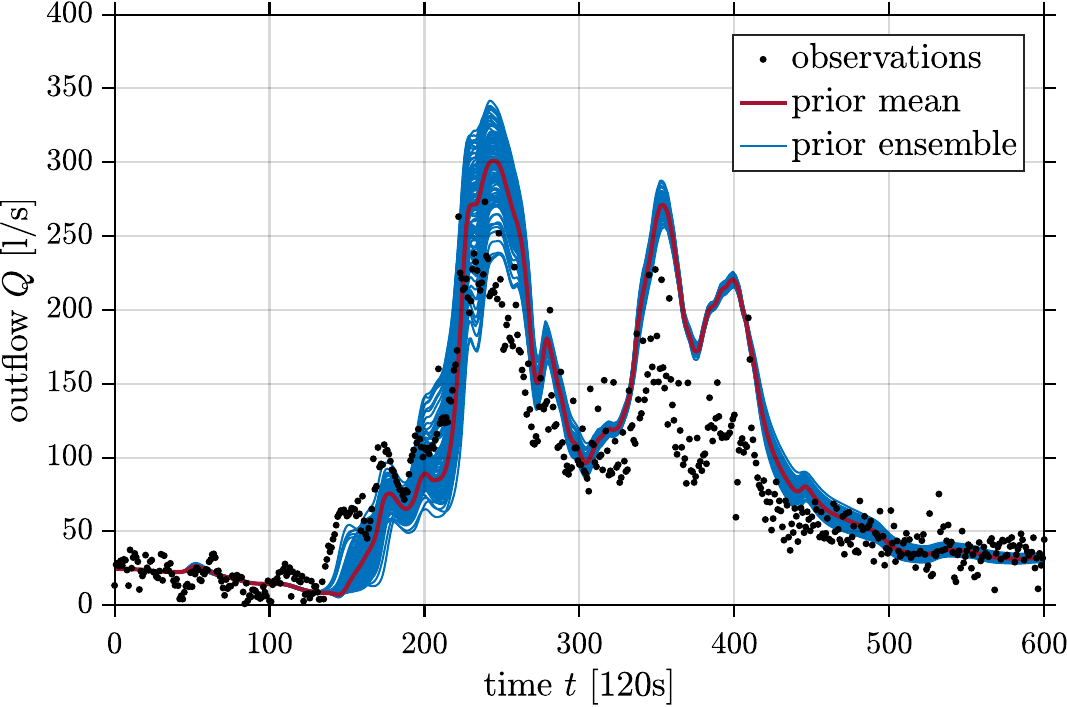}
    \subcaption{Prior predictions.}
    \label{fig:PCE:Prior}
  \end{minipage}\hfill
  \begin{minipage}[b]{\HYDROsubWidth}
    \centering
    \includegraphics[height=\HYDROfigHeight]{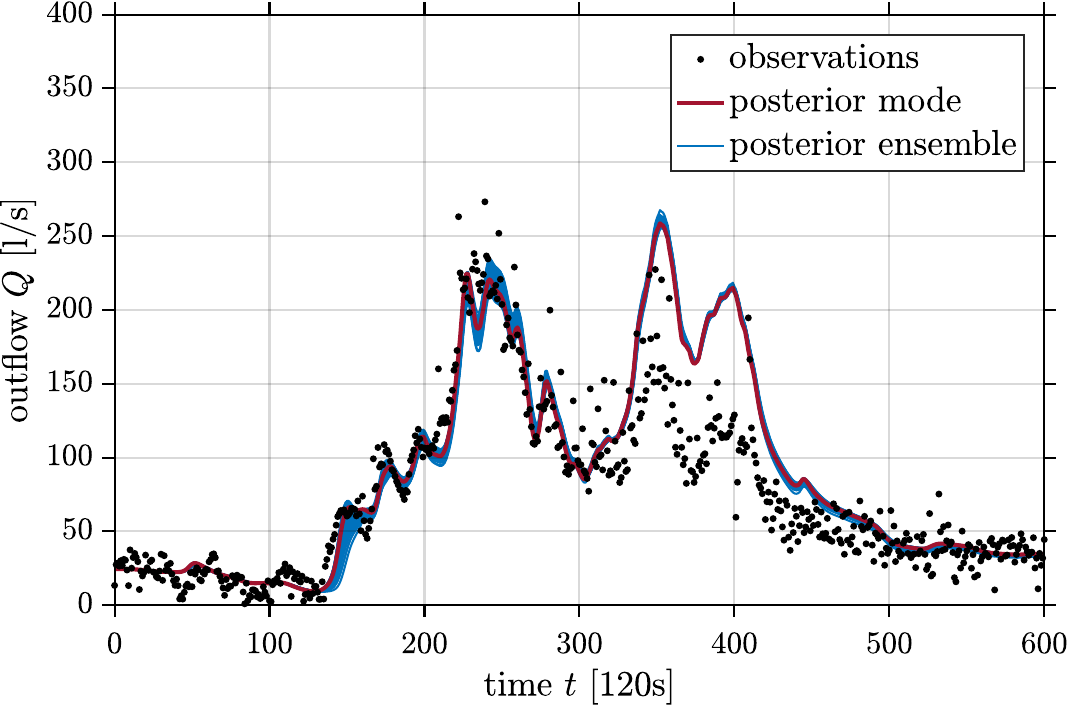}
    \subcaption{Posterior predictions.}
    \label{fig:PCE:Posterior} 
  \end{minipage}
  \caption{Stochastic model predictions.  Ensembles of predictions
    pertaining to the prior and posterior mode are shown in
    \cref{fig:PCE:Prior,fig:PCE:Posterior}, respectively.  As already
    seen in \cref{fig:Data:Outflow}, pronounced discrepancies between
    the predicted and observed outflow are found in the time intervals
    \([150,200]\) and \([250,500]\).}
  \label{fig:PCE} 
\setcounter{subfigure}{0}
\end{figure}

After having explored the posterior distribution, one can check the
obtained results for consistency by comparing an ensemble of prior and
posterior predictions with the data.  We start by comparing the
posterior \(\pi_1(\bm{x},\sigma \cond \bm{y})\) of the first model with
the correspondent prior \(\pi_1(\bm{x},\sigma)\) in this regard.  See
\cref{fig:PCE} for that purpose.  In \cref{fig:PCE:Prior}
the forecasts of the discharge are shown for one hundred input values
that were randomly sampled from the prior.  Likewise
\cref{fig:PCE:Posterior} shows the predictions for the same number of
posterior samples that were obtained from the MCMC chains by an
appropriate thinning.  Moreover, the time trajectory for the posterior
mode is highlighted.  The measurement uncertainty is not accounted for
in those figures.  As it can be seen, the prediction ensemble for the
prior contains more uncertainty than for the posterior.

The adjustment of the model parameters associated with the Bayesian
update does not significantly reduce the systematic discrepancy between
the simulated and the measured outflows from the drainage basin.  The
underlying reason is that varying the input parameters of the
hydrological simulator and the level of independent noise does not allow
for establishing full consistency between the simulations and the
observations, especially in the second half of the covered time
interval.  This was already clear after the discussion of
\cref{fig:Data:Outflow} and actually led to the inclusion of a
correlation and discrepancy term in the second model.

\begin{figure}[!ht]
    \centering
    \includegraphics[height=\HYDROfigHeight]{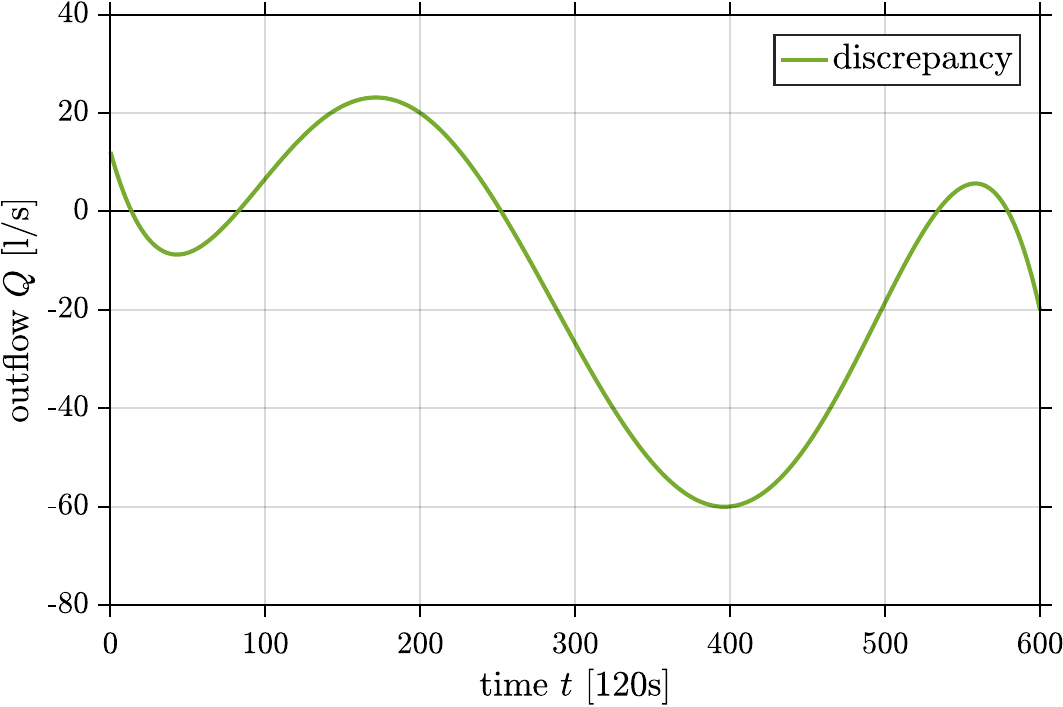}
    \caption{Model discrepancy.  The posterior mode reveals the bias
      of the simulator, especially in the time intervals \([150,200]\)
      and \([250,500]\), where mispredictions of the outflow occurred in
      \cref{fig:Data:Outflow}.}
    \label{fig:MAP:Discrepancy}
\end{figure}

\begin{figure}[!ht]
    \centering
    \includegraphics[height=\HYDROfigHeight]{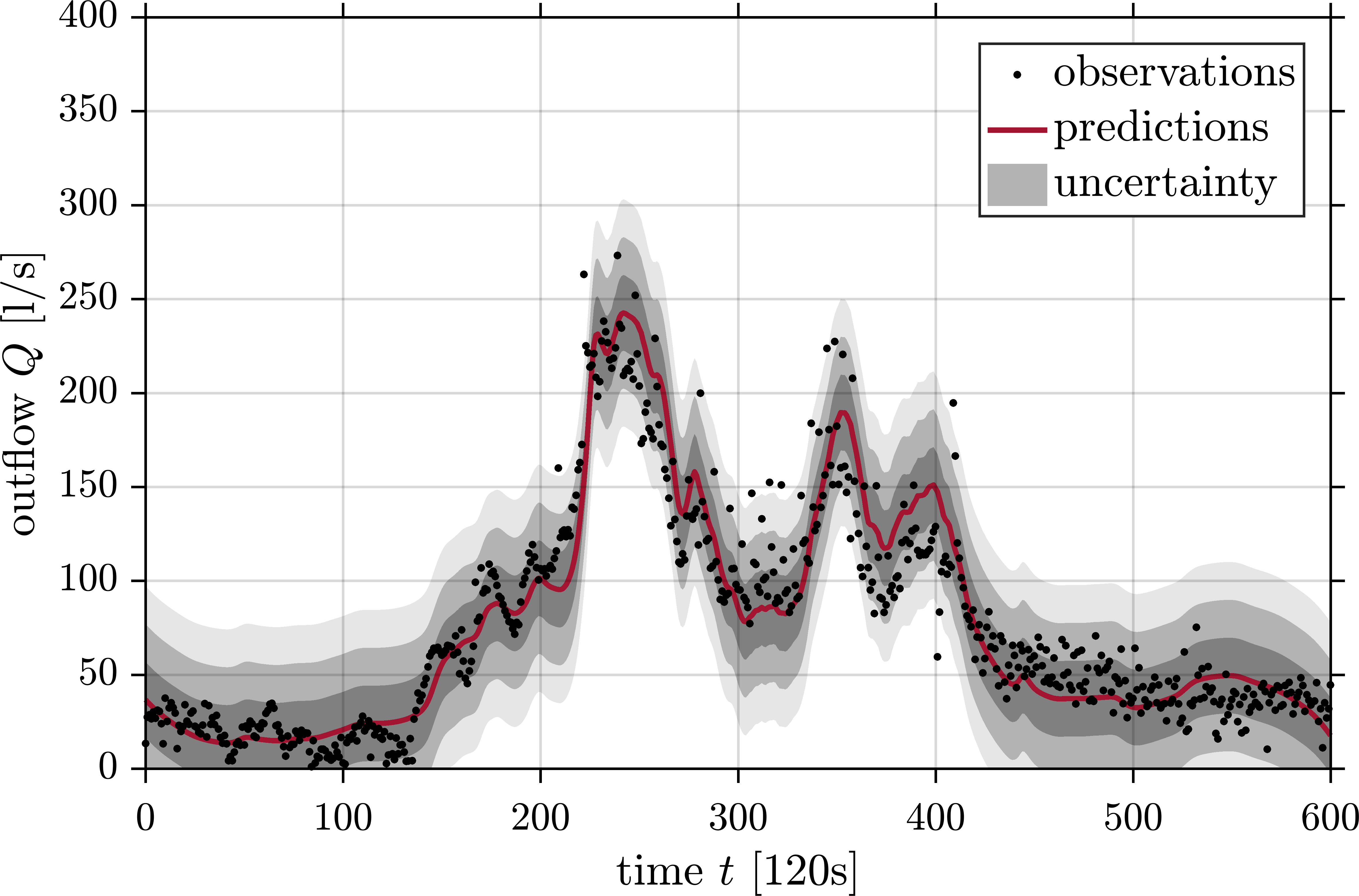}
    \caption{Corrected predictions.  Incorporating the bias identified
      in \cref{fig:MAP:Discrepancy} and the level of random errors
      corresponding to the posterior mode in \cref{fig:Post:Sigma} leads
      to well-calibrated predictions and uncertainty envelopes.}
    \label{fig:MAP:Predictions}
\end{figure}

We now investigate how well the posterior mode of
\(\pi_2(\bm{x},\bm{b},\sigma,\tau \cond \bm{y})\) aligns with the data.
The mode estimate of the discrepancy function \(\hat{\delta}(t) =
\delta(\hat{\bm{b}},t)\) is plotted in \cref{fig:MAP:Discrepancy}.  It
indicates a trend that the model underpredicts the actual rainfall in
roughly the interval \(t / \unit[120]{s} \in [100,250]\) and
overpredicts in \(t / \unit[120]{s} \in [250,500]\).  These
mis-predictions occur more or less for the period \(t / \unit[120]{s}
\in [100,450]\) of the precipitation event that was shown in
\cref{fig:Data:Rainfall}.
At the boundaries, say for \(t /
\unit[120]{s} \in [0,100]\) and \(t / \unit[120]{s} \in [500,600]\), the
discrepancy vanishes as far as the low-degree polynomial representation
admits.  The accordingly corrected predictions
\(\hat{\mathcal{M}}_p(\hat{\bm{x}}) + \hat{\bm{\delta}}\) are depicted
in \cref{fig:MAP:Predictions}.  They align with the data reasonably
well.  One, two and three \(\hat{\sigma}\) prediction intervals are
added so as to visualize the posterior mode prediction uncertainty.

\revision{
While the learned discrepancy term effectively captures the model errors as a function of time,
it is again noted that its specific form in \cref{fig:MAP:Discrepancy} pertains to the model and the studied rainfall event,
i.e.\ it does not generalize to unseen events.
The procedure, however, of separating and representing random noise and systematic error can be applied more generally.
}
\revision{It is further remarked that}, due to the additive and symmetric noise model,
the prediction intervals in \cref{fig:MAP:Predictions} extend to negative outflow values.
Since these values are physically nonsensical, they shall be ignored.

\section{Discussion and conclusion} \label{sec:Hydrology:DiscussionAndConclusion}

An efficient method for the non-intrusive emulation of time-dependent
hydrological systems was presented in this paper.  The idea was to
exploit typical characteristics of the system response based on
principal component analysis and sparse polynomial chaos expansions.
First of all, the dimensionality of the output space was reduced by
means of principal component analysis.  This takes advantage of a linear
correlation of the different response quantities.  Following this, the
reduced outputs were represented as polynomial chaos expansions.  This
enabled the utilization of sparsity patterns in the basis
decompositions.  All in all, a very fast and accurate emulator was
constructed this way.

The availability of a fast metamodel facilitated the probabilistic
uncertainty quantification of an urban drainage simulator.
Variance-based global sensitivity analysis and computational Bayesian
inference were accomplished for the dynamical simulation of a small
catchment area.  It was shown how the Sobol' indices of the predicted
outflow at various time instants can be efficiently computed based on
the polynomial expansions of the principal components.  The calibration
of the unknown hydrological parameters, random measurement uncertainties
and systematic model errors through Markov chain Monte Carlo sampling
was made possible.  In this context, a \revision{significant} speedup was achieved
through the substitution of the original simulator with the developed
surrogate model.

After all, it has to be said that the problem studied was rather simple.
While the catchment area under study had multiple outlets, only a single
one at the wastewater treatment plant was considered.  A very
coarse-grained parametrization of the unknowns based on crude averages
over hundreds of sub-catchments and channels was used.  Besides, the
whole case study was dependent on a single precipitation event for which
the rainfall record was taken as if it were measured without error.  It
is therefore envisaged to address more complex and realistic problems in
the future.  This may include problems that feature a finer-grained
representation of the spatially dependent input parameters and their
uncertainties, and models that explicitly acknowledge errors in the
rainfall input data.  This may also include the analysis of various
different precipitation events and longer time series data.  The
resulting increase of the input and output dimensionality can be coped
with on the basis of the proposed technique for structure-exploiting
surrogate modeling.

\section*{Software availability}

The urban drainage simulator was implemented with the storm water
management model (SWMM) \citep{Hydro:SWMM2015}.  For further details the
reader is redirected to the website
\href{https://www.epa.gov/water-research/storm-water-management-model-swmm}{https://www.epa.gov/water-research/storm-water-management-model-swmm}.
A metamodel based on principal component analysis and polynomial chaos
expansions was computed with UQLab
\citep{Computing:Marelli2014:Proc,Computing:Uqlab2015:Manual_1104}, a
Matlab-based framework for uncertainty quantification (UQ).  The
software is free for academic use and can be obtained from
\href{http://www.uqlab.com/}{http://www.uqlab.com/}.

\section*{Acknowledgements}
We thank the team of the COMCORDE project (SNF grant numbers
CR22I2~135551 and CR22I2~152824) for performing the rainfall-runoff
monitoring campaign and the many man-months spent on building the SWMM
model.  Moreover, we would like to thank David Machac for the execution
of the model training runs and Tobias Doppler for the excellent quality
control of the experimental data.  Juan Pablo Carbajal is acknowledged
for proofreading and commenting on the manuscript.

\appendix

\section{Principal component analysis} \label{sec:App:PCA}
Consider the random vector \(\tilde{\bm{Y}}\) with mean
\(\bm{\mu}_{\tilde{\bm{Y}}} = \mathds{E}[\tilde{\bm{Y}}]\) and
covariance matrix \(\bm{\Sigma}_{\tilde{\bm{Y}}} =
\mathrm{Cov}[\tilde{\bm{Y}}] =
\mathds{E}[(\tilde{\bm{Y}}-\bm{\mu}_{\tilde{\bm{Y}}})(\tilde{\bm{Y}}-\bm{\mu}_{\tilde{\bm{Y}}})^\top]\).
Since \(\bm{\Sigma}_{\tilde{\bm{Y}}}\) is symmetric and positive
definite, one can find linearly independent eigenvectors \(\bm{\phi}_i\)
with positive eigenvalues \(\lambda_i > 0\) for \(i =
0,\ldots,\dimData\).  The characteristic vectors and values satisfy
\begin{equation} \label{eq:Eigenequation}
  \bm{\Sigma}_{\tilde{\bm{Y}}} \bm{\phi}_i = \lambda_i \bm{\phi}_i.
\end{equation}
Eigenvectors corresponding to distinct eigenvalues are orthogonal
anyway, while they can be always chosen as such for degenerate
eigenvalues.  We assume that the eigenvalues are arranged in decreasing
order \(\lambda_0 \geq \lambda_1 \geq \ldots \geq \lambda_\dimData\) and
that eigenvectors are normalized such that \(\bm{\phi}_i^\top
\bm{\phi}_j = \delta_{ij}\) for \(i,j = 0,\ldots,\dimData\).  Leaving
degeneracy aside, this way the eigenvectors are uniquely defined up to a
multiplication by \(-1\).

The set of eigenvectors constitutes an orthonormal basis of
\(\mathds{R}^{\dimData+1} =
\mathrm{span}(\{\bm{\phi}_i\}_{i=0}^\dimData)\).  One can define the
orthogonal matrix \(\bm{\Phi} =
(\bm{\phi}_0,\bm{\phi}_1,\ldots,\bm{\phi}_\dimData)\) with
\(\bm{\Phi}^\top \bm{\Phi} = \bm{\Phi} \bm{\Phi}^\top = \bm{I}\).  It
diagonalizes the covariance matrix by
\begin{equation} \label{eq:Diagonaliztaion}
  \bm{\Phi}^\top \bm{\Sigma}_{\tilde{\bm{Y}}} \bm{\Phi}
  = \bm{\Lambda} = \begin{pmatrix}
                     \lambda_0 & 0          & \ldots & 0 \\
                     0         & \lambda_1  & \ldots & 0 \\
                     \vdots    & \vdots     & \ddots & \vdots \\
                     0         & 0          & \ldots & \lambda_\dimData \\
                   \end{pmatrix}.
\end{equation}
Vice versa, one obtains the spectral eigendecomposition of the covariance matrix
\(\bm{\Sigma}_{\tilde{\bm{Y}}} = \bm{\Phi} \bm{\Lambda} \bm{\Phi}^\top = \sum_{i=0}^\dimData \lambda_i \bm{\phi}_i \bm{\phi}_i^\top\).
Now consider the orthogonal transformation
\begin{equation} \label{eq:LinearTransformation}
  \tilde{\bm{Z}} = \bm{\Phi}^\top (\tilde{\bm{Y}} - \bm{\mu}_{\tilde{\bm{Y}}}).
\end{equation}
The linearly transformed random vector has mean zero \(\mathds{E}[\tilde{\bm{Z}}] = \bm{0}\)
and the diagonal covariance matrix \(\mathrm{Cov}[\tilde{\bm{Z}}] = \mathds{E}[\tilde{\bm{Z}} \tilde{\bm{Z}}^\top] = \bm{\Lambda}\), i.e.\ it has been centered and decorrelated.
Independence is not necessarily implied thereby.
Though, the special case involving Gaussianity forms an exception.
The back-transformation reads
\begin{equation} \label{eq:BackTransformation}
  \tilde{\bm{Y}} = \bm{\mu}_{\tilde{\bm{Y}}} + \bm{\Phi} \tilde{\bm{Z}} = \bm{\mu}_{\tilde{\bm{Y}}} + \sum\limits_{i=0}^\dimData \tilde{Z}_i \bm{\phi}_i.
\end{equation}
This is the discrete KL expansion of the random vector \(\tilde{\bm{Y}}\).
The random variables \(\tilde{Z}_i = \bm{\phi}_i^\top (\tilde{\bm{Y}}-\bm{\mu}_{\tilde{\bm{Y}}})\) for \(i = 0,\ldots,\dimData\) are called the \emph{principal components}.

Define the \emph{total variance} of \(\tilde{\bm{Y}}\) as the sum \(\sum_{i=0}^\dimData \mathrm{Var}[\tilde{Y}_i]\) of the individual variances of \(\tilde{Y}_i\).
The orthogonal transformation preserves the total variance in the sense that
\begin{equation} \label{eq:TotalVariance}
  \sum_{i=0}^\dimData \mathrm{Var}[\tilde{Y}_i] = \operatorname{tr}(\bm{\Sigma}_{\tilde{\bm{Y}}})
  = \operatorname{tr}(\bm{\Lambda}) = \sum_{i=0}^\dimData \mathrm{Var}[\tilde{Z}_i] = \sum_{i=0}^\dimData \lambda_i.
\end{equation}
This follows from the invariance of the trace under cyclic permutations.
The KL expansion is optimal with respect to compaction of the total
variance.  Consider keeping only the first \(\dimData^\prime + 1 \leq
\dimData + 1\) terms in
\begin{equation} \label{eq:TruncatedKLE}
  \tilde{\bm{Y}} \approx \bm{\mu}_{\tilde{\bm{Y}}} + \sum\limits_{i=0}^{\dimData^\prime} \tilde{Z}_i \bm{\phi}_i.
\end{equation}
This is the expansion that contains most of the total variance with
\(\dimData^\prime + 1\) terms only.  The number of terms is normally
chosen such that at least a predetermined fraction
\(\sum_{i=0}^{\dimData^\prime} \lambda_i / \sum_{i=0}^\dimData
\lambda_i\) of the total variance is explained.

\bibliography{bib}

\end{document}